\documentclass{aa}  
\usepackage{pst-all}
\usepackage{txfonts,makecell}
\usepackage{amsmath,amssymb}
\usepackage{graphicx}
\usepackage{array,multirow}
\usepackage[alsoload=astronomy]{siunitx}
\DeclareSIUnit\ergon{erg}
\bibpunct{(}{)}{;}{a}{}{,} 
\usepackage[pdfauthor={Barbosa et al.},
            pdftitle={Sloshing in its cD halo: MUSE kinematics of the central galaxy NGC 3311}]{hyperref}
\hypersetup{colorlinks=true, linkcolor=blue, citecolor=blue}
\usepackage{acronym}

\acrodef{BCG}[BCG]{Brightest Cluster Galaxy}
\acrodef{BPT}{Baldwin-Phillips-Terlevich}
\acrodef{CFHT}{Canada France Havai Telescope}
\acrodef{ETGs}{Early-Type Galaxies}
\acrodef{FMD}{Finite Mixture Distribution} 
\acrodef{FORS2}{Focal Reducer/ Low Dispersion Spectrograph 2}
\acrodef{FOV}{field-of-view}
\acrodef{FWHM}{Full Width at Half Maximum}
\acrodef{HST}{Hubble Space Telescope}
\acrodef{ICL}{Intra Cluster Light}
\acrodef{ISM}{Interstellar Medium}
\acrodef{IFU}{Integral Field Unit}
\acrodef{IMF}{Initial Mass Function}
\acrodef{KDC}{Kinematically Decoupled Core}
\acrodef{LOS}[LOS]{line-of-sight}
\acrodef{MCMC}{Monte Carlo Markov Chain}
\acrodef{MUSE}[MUSE]{Multi Unit Spectroscopic Explorer}
\acrodef{LOSVD}[LOSVD]{Line-of-sight Velocity Distribution}
\acrodef{PSF}{Point Spread Function}
\acrodef{SDSS}{Sloan Digital Sky Survey}
\acrodef{S/N}[S/N]{Signal-to-Noise Ratio}
\acrodef{SSP}[SSP]{Single Stellar Population}
\acrodef{VLT}{Very Large Telescope}

\title{Sloshing in its cD halo: MUSE kinematics of the central galaxy NGC 3311 in the Hydra I cluster}        
\author{C. E. Barbosa\inst{,\ref{usp}}\fnmsep\thanks{Corresponding author: {\tt carlos.barbosa@usp.br}} \and
              M. Arnaboldi\inst{\ref{eso},\ref{inaf}} \and
              L. Coccato\inst{\ref{eso}} \and
              O. Gerhard\inst{\ref{mpe}} \and
              C. Mendes de Oliveira\inst{\ref{usp}} \and
              M. Hilker\inst{\ref{eso}} \and
              T. Richtler\inst{\ref{concepcion}}
              }

\institute{
Universidade de S\~ao Paulo, IAG, Departamento de Astronomia, Rua do Mat\~ao 1226, S\~ao Paulo, SP, Brazil\label{usp} 
\and
European Southern Observatory,  Karl-Schwarzschild-Stra\ss{}e 2, 85748, Garching, Germany\label{eso} 
\and
INAF, Osservatorio Astronomico di Torino, STrada Osservatorio 20, 10025 Pino Torinese, Italy\label{inaf}
\and
Max-Planck-Institut fur Extraterrestrische Physik, Giessenbachstrasse, 85741 Garching, Germany\label{mpe} 
\and
Departamento de Astronomia, Universidad de Concepci\'on, Concepci\'on, Chile\label{concepcion}     
}

   \date{Received August 25, 2017; accepted}

  \abstract
   {Early-type galaxies (ETGs) show a strong size evolution with redshift. This evolution is explained by fast ``in-situ'' star formation at high-$z$ followed by a late mass assembly mostly driven by minor mergers that deposit stars primarily in the outer halo.}
   {We aim to identify the main structural components of the Hydra~I cD galaxy NGC~3311 to investigate the connection between the central galaxy and the surrounding stellar halo.}
   {We produce maps of the line-of-sight velocity distribution (LOSVD) moments from a mosaic of MUSE pointings covering NGC~3311 out to $25$ kpc. Combining deep photometric and spectroscopic data, we model the LOSVD maps using a finite mixture distribution, including four non-concentric components that are nearly isothermal spheroids, with different line-of-sight systemic velocities $V$, velocity dispersions $\sigma$, and small (constant) values of the higher order Gauss-Hermite moments $h_3$ and $h_4$. }
   {The kinemetry analysis indicates that NGC~3311 is classified as a slow rotator, although the galaxy shows a line-of-sight velocity gradient along the photometric major axis. The comparison of the correlations between $h_3$ and $h_4$ with $V/\sigma$ with simulated galaxies indicates that NGC~3311 assembled mainly through dry mergers. The $\sigma$ profile rises to $\simeq 400$ km s$^{\text -1}$ at 20 kpc, a significant fraction (0.55) of the Hydra~I cluster velocity dispersion, indicating that stars there were stripped from progenitors orbiting in the cluster core. The finite mixture distribution modeling supports three inner components related to the central galaxy and a fourth component with large effective radius ($51$ kpc) and velocity dispersion ($327$ km s$^{\text{-1}}$) consistent with a cD envelope. We find that the cD envelope is offset from the center of NGC~3311 both spatially (8.6 kpc) and in velocity ($\Delta V = 204$ kms$^{-1}$), but coincide with the cluster core X-ray isophotes and the mean velocity of core galaxies. Also, the envelope contributes to the broad wings of the LOSVD measured by large $h_4$ values within 10 kpc.}
{The cD envelope of NGC~3311 is dynamically associated with the cluster core, which in Hydra~I is in addition displaced from the cluster center, presumably due to a recent subcluster merger.}
   

\begin{document}

   \maketitle
%

\section{Introduction}
\label{sec:intro}

The nearly featureless morphologies of massive \ac{ETGs} and their radial surface brightness profiles have been a test benchmark for any theories of galaxy formation and evolution since the introduction of the $R^{1/4}$ law \citep{1953MNRAS.113..134D}.  This simple analytical formula captures the high degree of central concentration of the light in these objects, but also the large radial extension of their light distribution \citep{2013ApJ...766...47H}. At the brightest end, the most luminous galaxies in the early-type family, the cD galaxies, are embedded in extended envelopes \citep{1986ApJS...60..603S}, with shallow surface brightness gradient profiles thus fading into the cluster of galaxies that surrounds them.

Recent observations show that massive, passively evolving galaxies are identified already at $z\sim 2.5$ \citep{2004Natur.430..184C}. In particular, galaxies in the ``red nuggets'' population are considered to be precursors of the nearby giant \ac{ETGs} \citep{2009Natur.460..717V,2010ApJ...714L..79C,2015ApJ...813...23V}, although with smaller sizes by a factor $\sim 3$ and higher central stellar velocity dispersions ($\sigma$) than the local \ac{ETGs} with similar stellar mass. Among different proposed formation models, the two-phase formation scenario \citep{2007MNRAS.375....2D,2010ApJ...725.2312O,2012ApJ...744...63O} is able to satisfy these observational constraints. In this model, the central region of \ac{ETGs} are formed in fast dissipative process early in the history of the universe ($z\geq 3$), while the evolution at low redshift is dominated by the assembly of a stellar halo, which is stochastically accreted as a consequence of mostly dry mergers \citep[see also][]{2013MNRAS.434.3348C}. Such a picture can be probed at low redshift by the observations of cD galaxies, where the processes of late mass accretion are believed to be extreme owing to the large occurrence of galaxy disruption processes that are inherent to the strong gravitational interactions within their massive dark halos. 

Currently, several avenues are being pursued to identify the different signatures left by the dissipative (also dubbed {\it in situ}) and accretion (also dubbed {\it ex-situ}) processes in massive \ac{ETGs}. For instance, the radial gradients of the stellar populations (metallicity, age, abundance ratios) of \ac{ETGs} preserve information about accretion history \citep[e.g.,][]{2015MNRAS.449..528H,2016ApJ...833..158C}, and some attempts to determine gradients out to large radius were carried out using long-slit \citep{2010MNRAS.407L..26C,2011A&A...533A.138C}, multi-slit \citet{2016A&A...589A.139B} and \ac{IFU} \citep{2013ApJ...776...64G,2015ApJ...807...11G} spectroscopy.  Alternatively, information about the accretion events may be present in the surface brightness profiles.  See the recent results in \citet{2017arXiv170310835S} for decompositions that are motivated by the predictions of cosmological simulations, and \citet{2013ApJ...766...47H} and \citet{2005ApJ...618..195G} for the identification of galaxy and halo components in extended 2D photometric data. 

One open question is whether the 2D surface brightness decomposition in multiple components is supported independently by the maps of the \ac{LOSVD} moments, e.g., whether these photometric components can be identified in the kinematic profiles also. Differently from late-type galaxies, where the bulge/disk decomposition of photometric and kinematic profiles is performed routinely \citep[e.g.,][]{2011MNRAS.414..642C}, the identification of the kinematic signatures of the inner and outer stellar components of \ac{ETGs} are  hampered primarily by the low \ac{S/N} in the spectroscopic absorption line measurements, expecially  at large galactocentric distances. 

Recently, \citet{2015ApJ...807...56B} dealt with the above question in cD galaxies, in the study of NGC~6166. They used a combination of deep long-slit observations with deep photometry and required that the photometric decomposition reproduced the radial gradients of the four LOSVD moments, i.e. $V$, $\sigma$, $h_3$, $h_4$. As a result, the best kinematic \& photometric model identified two spheroids with low S\'ersic indices ($n<4$), with very different surface brightnesses and radii, and also different systemic velocities and velocity dispersions. Because of the large value of its effective radius, the outer component was identified with the cD envelope of NGC~6166, and its large velocity dispersion supports the idea that its stars were stripped from  galaxies orbiting in Abell 2199. These results highlight the importance of substructures to understand the 2D photometric and kinematic maps of ETGs, but they also raise interesting questions.  Are these components useful parameterisations only, or do they comply with the scaling relations for galaxies, like the Faber-Jackson \citep{1976ApJ...204..668F} or the Fundamental plane \citep{1973A&A....23..259B,1987ApJ...313...59D,1987ApJ...313...42D} relations? Do they qualify as physically distinct components in these galaxies, like bulges and disks in spirals? 

In this work, we study the cD galaxy NGC~3311, located at the core of the Hydra~I (Abell 1060) cluster, one of the nearest prototypical massive elliptical galaxies. Photometrically, NGC~3311 is the \ac{BCG} of the cluster, and has an extended and diffuse stellar halo on top of a high surface brightness central component. As shown by \citet{2012A&A...545A..37A}, the surface brightness profile of NGC~3311 requires more than one component to describe its overall asymmetry with respect to the galaxy's luminous center. Long-slit observations indicate that NGC~3311 has a velocity dispersion profile that rises from a central value of $\sigma_0\approx$\SI{175}{\kilo\meter\per\second} to $\sigma\approx$\SI{400}{\kilo\meter\per\second} at major axis distances of $20$ kpc \citep{2010A&A...520L...9V,2011A&A...531A.119R}. Moreover, features in the velocity dispersion profile are correlated with photometric \citep{2015IAUS..309..221H,hilker2017} and stellar population substructures \citep{2011A&A...533A.138C,2016A&A...589A.139B}. These studies indicate that NGC~3311 may also be described as a central galaxy surrounded by a cD envelope, similarly to NGC~6166,.

Our goal is to derive a coherent description of NGC 3311 that can account for both the surface brightness and the \ac{LOSVD}, to learn about its formation processes. Expanding on the approach used by \citet{2015ApJ...807...56B}, we perform a two-dimensional modeling of the data by matching simultaneously 2D surface brightness and  \ac{LOSVD} moments maps. For this purpose, we use recently acquired new deep integral field observations for the central region of the Hydra I cluster with the \ac{MUSE} IFU, that allows a detailed study of the kinematics of NGC~3311 with unprecedented spatial resolution, together with deep V-band archival data. 

We structure this work as follows. In Section~\ref{sec:kinematics}, we present the analysis of the \ac{MUSE} data set, including the mosaicing strategy of the observations, data reduction, methodology for \ac{LOSVD} measurement and validation of the results, and the analysis of the kinematics. In a companion paper \citep{hilker2017}, we also explore the kinematics of NGC~3311 at larger radii using multiple slit mask spectroscopy with FORS2 on VLT.

In Section~\ref{sec:photometry}, we perform a detailed photometric decomposition of NGC~3311 surface brightness with multiple spheroidal components. In Section~\ref{sec:modeling}, we carry out a unified model of both photometric and kinematic data using the method of finite mixture distribution. In Section~\ref{sec:discussion}, we discuss the implications of our results for the understanding of cD galaxies, the peculiar velocity of inner galaxy and envelope, the origin of their velocity and radial biases, and the strong radial variation of the LOSVD moments. We summarize and conclude our work in Section~\ref{sec:conclusion}. 

Throughout this work, we adopt the distance to the Hydra~I cluster of $D=50.7$ Mpc, based on the Hubble flow with $H_0=70.5$ \si{\kilo\meter\per\second\per\mega\parsec} \citep{2009ApJS..180..330K} assuming a radial velocity of \SI{3777}{\kilo\meter\per\second} \citep{1999ApJS..125...35S}, which results in $1''=0.262$ kpc. We assume an effective radius of $R_e=8.4$ kpc for NGC~3311 \citep{2012A&A...545A..37A}. 

\section{Kinematic analysis of MUSE observations}
\label{sec:kinematics}

\subsection{Observations and data reduction}
\label{sec:reduction}

We carried out integral field observations of the core of the Hydra~I cluster using the \ac{MUSE} instrument \citep{2003SPIE.4841.1096H,2004SPIE.5492.1145B}, mounted at the Nasmyth focus of the UT4 8m telescope of the \ac{VLT}, under ESO programme 094.B-0711A (PI: Arnaboldi). Observations were taken in the wide field mode, that covers a $1\times 1$ arcmin$^{\rm 2}$ \ac{FOV} with spatial sampling of $0.2\times 0.2$ arcsec$^{\rm 2}$. In the case of NGC~3311, it corresponds to a region of $15.7\times 15.7$ kpc$^{\rm 2}$. The wavelength coverage is $4650\leq\lambda($\r{A}$)\leq 9300$ with resolving power varying between 2000 and 4000. 

\begin{figure}
 \centering
 \includegraphics[width=0.98\linewidth]{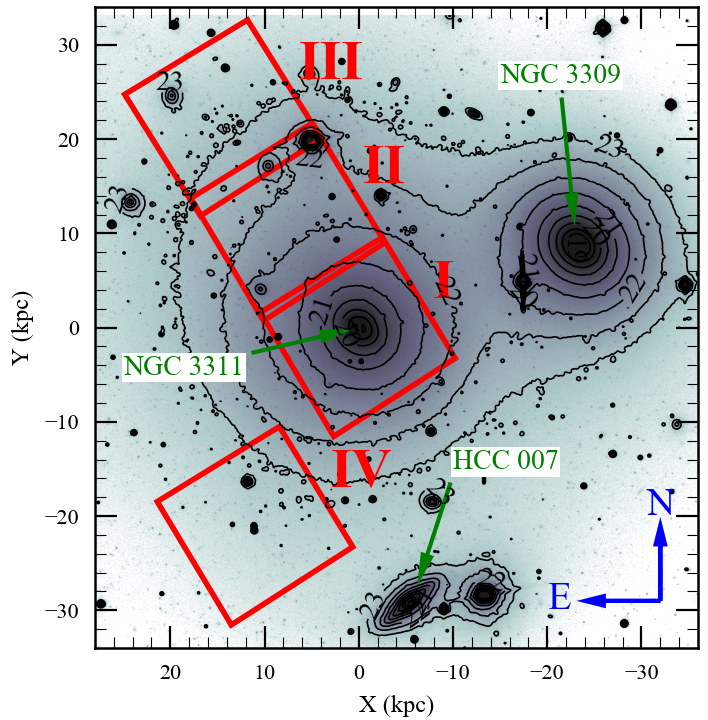}
 \caption[Area coverage of the MUSE observations of NGC~3311, in the core of at the Hydra~I cluster.]{Area coverage of the MUSE observations of NGC~3311, in the core of the Hydra~I cluster. The mosaicing consists of four fields; three are aligned with the photometric major axis (I, II, and III) and survey the central region of the galaxy and the NE substructure. The fourth pointing (field IV) is placed on the extended diffuse tail off the lenticular galaxy HCC~007. The V-band image from \citet{2012A&A...545A..37A} is used in the background, and the black lines show the V-band contours in the range from 21 to 23.5 mag arcsec$^{\rm -2}$ in intervals of 0.5 mag arcsec$^{\rm -2}$.}
 \label{fig:strategy}
 \end{figure}

Fig.~\ref{fig:strategy} shows the mosaic of pointings adopted in our observations, including four fields, I, II, III and IV, with exposure times of \SI{730}{\second}, \SI{670}{\second}, \SI{1015}{\second} and \SI{1370}{\second} respectively. The first three fields are positioned along the major axis of NGC~3311 to probe the center of NGC~3311 and the NE substructure observed in X-ray \citep{2004PASJ...56..743H} and optical \citep{2012A&A...545A..37A} imaging, as well as in the metallicity map \citep{2016A&A...589A.139B}. We have an additional pointing, field IV, located next to HCC~007, a spectroscopically confirmed member of the cluster with a systemic velocity of $V=4830\pm13$ \si{\kilo\meter\per\second} \citep{2008A&A...486..697M}, that is intended to cover the diffuse stellar tidal tail observed by \citet{2012A&A...545A..37A}. The large companion NGC 3309 is also bound to the Hydra I cluster, with a central systemic velocity of $V=4099 \pm 27$ \si{\kilo\meter\per\second} \citep{2008A&A...486..697M}. However, NGC 3309 does not seem to be interacting with NGC 3311, as indicated by the lack of distortions in the photometric profiles of both galaxies, and contributes to the surface brightness profile of NGC 3311 only at distances larger than 20 kpc \citep[see also Section \ref{sec:sercomponents}]{2012A&A...545A..37A}. 

The data reduction was executed with the MUSE pipeline \citep{2012SPIE.8451E..0BW}, run under the Esoreflex environment \citep{2013A&A...559A..96F}, that provides real-time visualization of the data processing. We used the standard recipes for the instrument for the reduction, which includes flat fielding, bias correction, wavelength calibration, sky subtraction, flux calibration and combination of the cubes. We masked out from the analysis only the saturated star observed in the upper region of field II and its diffraction spikes that are also observed in field III.

In this work, we are interested in the analysis of the \ac{LOSVD} of the stars in the stellar halo of NGC 3311 only.
The large population of globular low-mass systems in the central region of the Hydra I cluster core, including dwarf galaxies \citep{2008A&A...486..697M} and ultra-compact dwarfs \citep{2011A&A...531A...4M}, will be the subject of a forthcoming paper, and are masked in the current analysis. The masking of these sources is carried out on the white lamp images using the segmentation image produced by \textsc{Sextractor} \citep{1996A&AS..117..393B} using a $1.5\sigma$ threshold above the local continuum. 

\subsection{Determination of the stellar line-of-sight velocity distributions}
\label{sec:kinmethod}

We extracted the kinematics with full spectrum fitting using the penalized pixel-fitting (\textsc{pPXF}) program \citep{2004PASP..116..138C}, which models the observations by a combination of spectral templates convolved with a \ac{LOSVD} parametrized as a Gauss-Hermite profile \citep[see][]{1993MNRAS.265..213G,1993ApJ...407..525V}. A review of \textsc{pPXF} is presented in \citet{2017MNRAS.466..798C}. In the following, we summarize the steps performed in our analysis.

\begin{figure}
\centering
\includegraphics[width=0.99\linewidth]{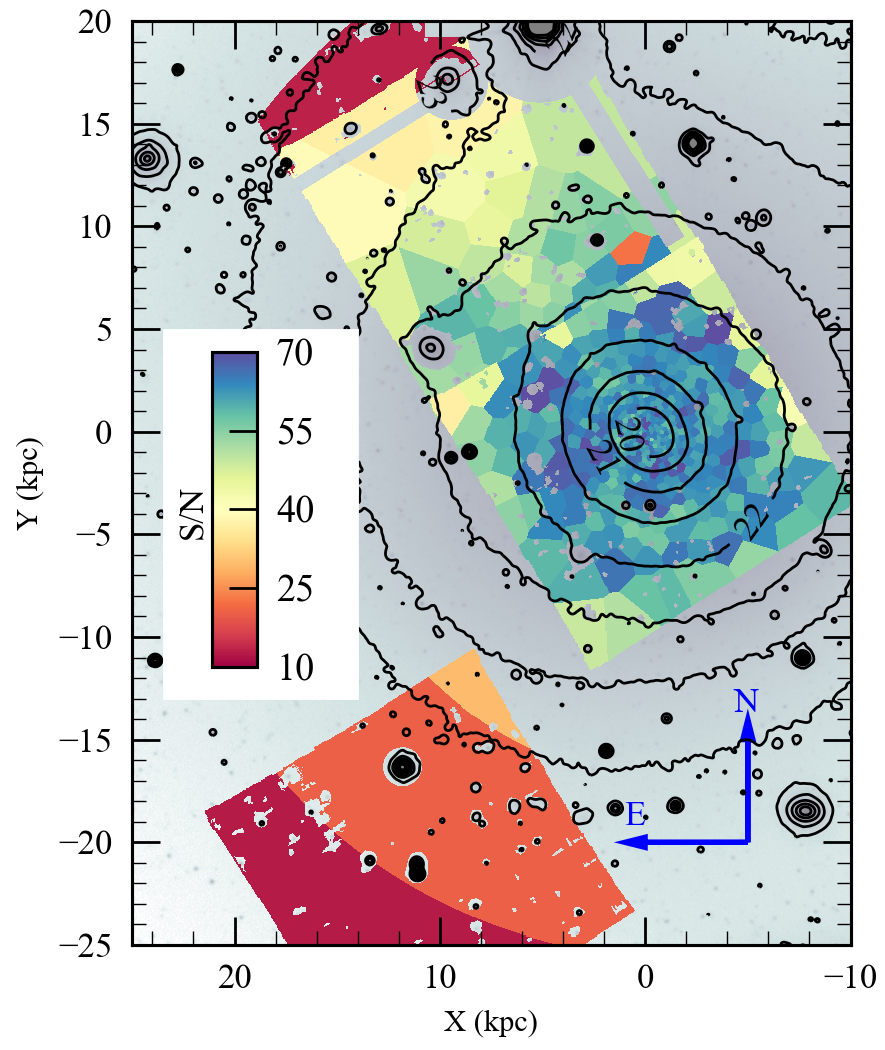}
\caption[Binning scheme and resulting S/N distribution of the spectra over the stellar halo in NGC ~3311.]{Binning scheme and resulting S/N distribution of the spectra over the stellar halo in NGC~3311 from the MUSE IFU data. Spatial bins are regions of constant S/N and color. Gray regions within the MUSE fields indicate masked spaxels, including low signal to noise (S/N $<10$) bins, saturated stars and its diffraction spikes, and all objects detected with \textsc{sextractor} which may contaminate the measurement of the LOSVDs of the stellar halo.}
\label{fig:sn}
\end{figure}

Considering that we are interested in the study of the third ($h_3)$ and fourth ($h_4$) moments of the \ac{LOSVD}, we combined spectra from adjacent unmasked spaxels to increase the \ac{S/N} of the data with the Voronoi tesselation from \citet{2003MNRAS.342..345C}. We used the target signal-to-noise of $\text{S/N}\approx 70$ which, according to simulations from \citet{2004PASP..116..138C}, allows the recovery of the \ac{LOSVD} with a precision better than  5\% in the regime of the velocity dispersions found in NGC~3311, i.e., $\sigma>150$ \si{\kilo\meter\per\second}. However, this large \ac{S/N} would require the combination of the entire \ac{FOV} in fields III and IV, which would completely erase any spatial information. Therefore, in addition to the Voronoi binning, we also separated the data into a few radial bins. Fig.~\ref{fig:sn} illustrates the resulting binning scheme and the \ac{S/N} obtained in this combination process.

We fitted each spectrum with two kinematic components simultaneously, one for the stellar component, that is our main focus in this work, and another component for the gas emission, that is used to improve the fitting. The stellar \ac{LOSVD}s are calculated using \ac{SSP} templates from \citet{2010MNRAS.404.1639V}, constructed with stellar spectra of the MILES library \citep{2006MNRAS.371..703S,2011A&A...532A..95F}, considering a bimodal \acused{IMF} \acl{IMF} \citep[\acs{IMF}, see][]{1996ApJS..106..307V}, metallicities in the range $-0.66\leq$[Z/H]$\leq 0.40$, ages between 0.1 and 14 Gyr and two different alpha-element abundances (0 and 0.4). The gas \ac{LOSVD}s are computed using a set of Gaussian emission line templates including H$\beta$ ($\lambda 4861$), [O III] ($\lambda 4959$, $\lambda 5007$), H$\alpha$ ($\lambda 6565$), [N II] ($\lambda 6585$, $\lambda 6550$) and [S II] ($\lambda 6718$, $\lambda 6733$). Considering that the \ac{MUSE} spectra has a resolution that varies as a function of the wavelength, we homogenized the spectral resolution of the observations and templates to obtain a \ac{FWHM} of 2.9 \r{A}.

We used additive polynomials to correct for small variations in the flux calibration between the observations and templates. After the visual inspection of the results, we noticed that our initial constraints for the velocity dispersion in \textsc{pPXF} for the gas component ($\sigma<1000$ \si{\kilo\meter\per\second}) resulted in poor fittings because of template mismatch. To avoid such cases, we additionally constrain the fitting by assuming  $\sigma_{\rm gas}<80$ \si{\kilo\meter\per\second}, which is approximately the velocity dispersion of emission lines in the central kiloparsec of NGC 3311, considering that we have not observed any occurrence of strong, broad emission lines.  

\begin{figure}
\centering
\includegraphics[width=0.98\linewidth]{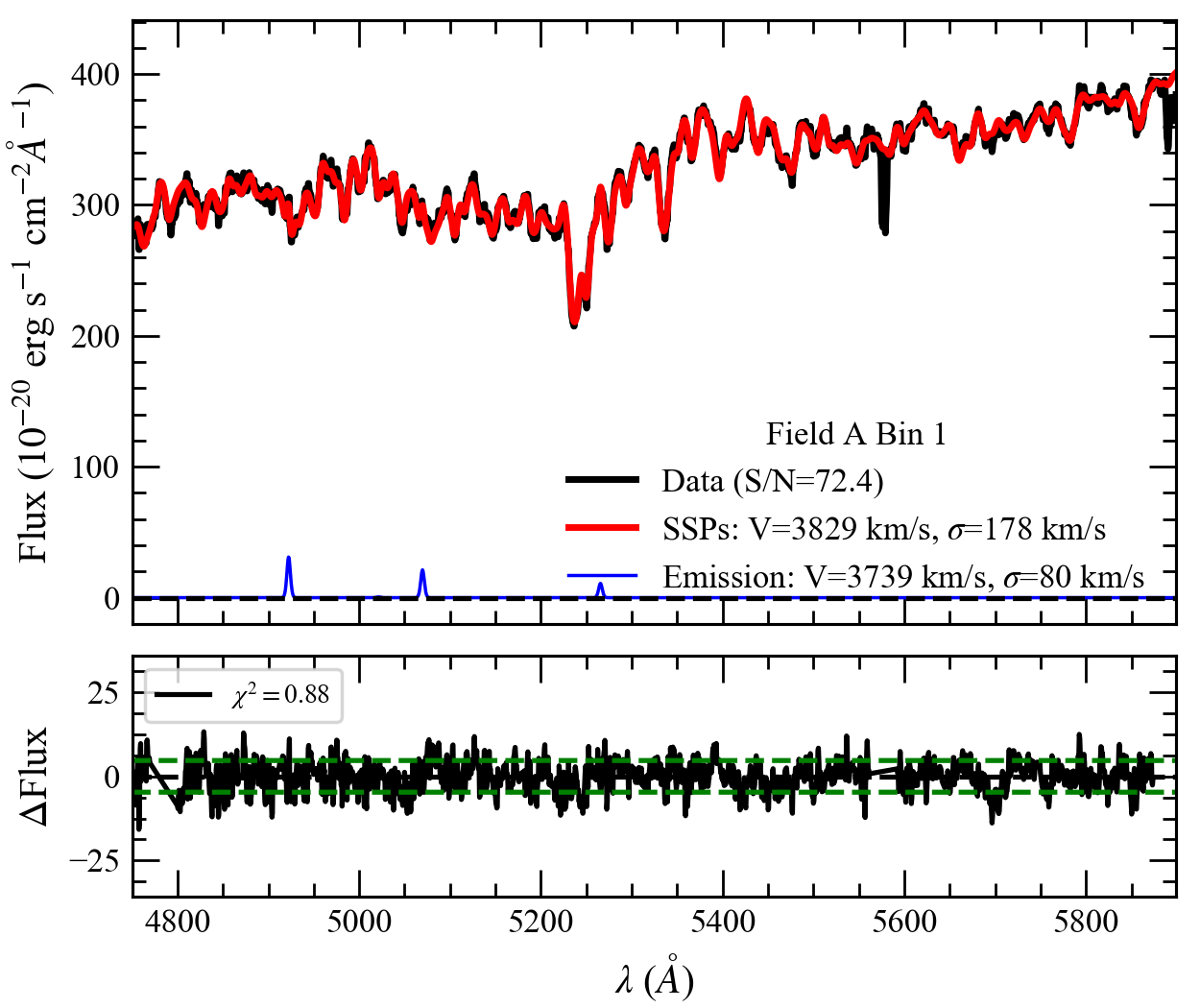}
\caption{Example of the fitting process with \textsc{pPXF} for the MUSE observations of the Hydra~I cluster core. The upper panel displays the observed spectrum (black), the best fit for the stellar component (red) and the best fit for the emission lines (blue). The lower panel shows the residuals of the fitting (solid black) and the standard deviation of the noise (dashed green).}
\label{fig:example}
\end{figure}

The wavelength range of the fitting was shown to be of importance in our analysis. The combination of the MUSE observations and the MILES templates ranges allows the determination of the \ac{LOSVD} using a large wavelength range between $4700\lesssim \lambda ($\r{A}$)\lesssim 7500$.  We tested the effect of using different wavelength ranges within this dominion, to check for consistency, and notice that the central velocity dispersion may increase by $\sim 30$ \si{\kilo\meter\per\second} depending on the inclusion wavelengths greater than $6000$ \r{A}. This effect is more critical in the center of NGC~3311, where emission lines are stronger, and where \ac{ISM} absorption lines can cause systematic effects in the derived \ac{LOSVD}s. A more detailed description of the effects of faint emission lines for the determination of the \ac{LOSVD} is presented in Appendix~\ref{sec:ism}. For simplicity and consistency, we restricted our analysis to wavelengths smaller than $5900$ \r{A}. Regions with strong residual sky lines, such as those found at $\lambda=4785$ \r{A}, $\lambda=5577$ \r{A}, and $\lambda=5889$ \r{A} are excluded from the fitting process. Fig.~\ref{fig:example} illustrates the results of our fitting for the central bin of NGC~3311. All velocities were corrected to the heliocentric velocity, calculated with the \textsc{IRAF} task \textsc{rvcorrect}. 

\begin{figure}
\centering
\includegraphics[width=\linewidth]{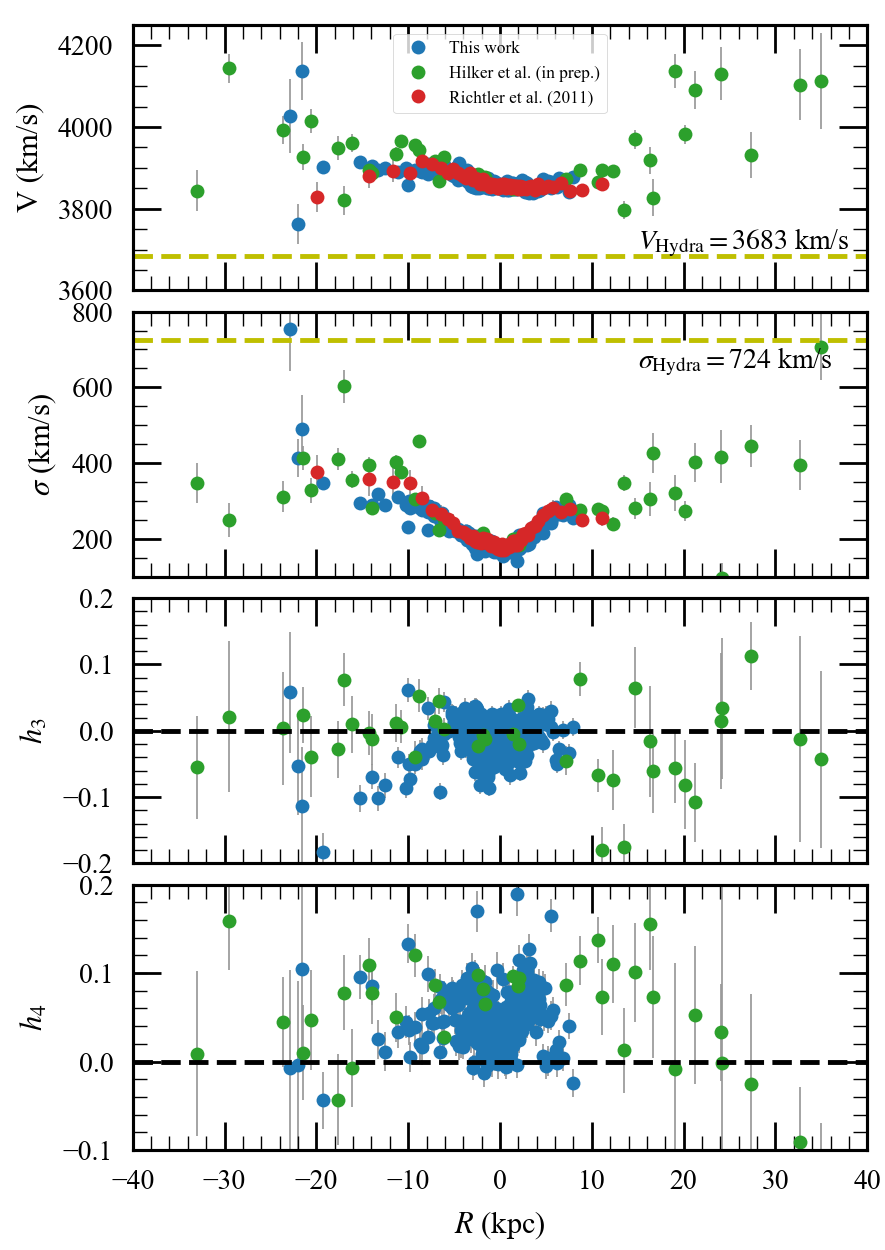}
\caption{Comparison of the profiles of the four moments of the LOSVDs from this work with literature data. Red circles show the results from long-slit observations of \citet{2011A&A...531A.119R} at the position angle of \SI{40}{\degree} and centered on NGC~3311. Blue (green) circles show the results from this work \citep{hilker2017} extracted along the same position angle within a pseudo-slit of \SI{20}{\arcsec} (\SI{40}{\arcsec}) width. The dashed yellow line in the first and second panel indicate the cluster's velocity and velocity dispersion according to \citet{2003ApJ...591..764C}. Positive (negative) radius indicate data points towards the southeast (northwest) direction from the center of the galaxy.} 
\label{fig:kinmajaxis}
\end{figure}

To test our method and have an initial evaluation of the results, we perform a comparison of our results with the literature in Fig.~\ref{fig:kinmajaxis}, where we show the radial profiles of the four moments of the \ac{LOSVD}s along the major axis of NGC 3311 at a position angle of \SI{40}{\degree}. To separate the results from the two sides of the galaxy, we folded the radial axis around the center adopting negative and positive radius for the data towards the northeast and southwest directions respectively. The upper two panels display the velocity and velocity dispersion from \citet{2011A&A...531A.119R}, who observed NGC~3311 using long-slit observations. We also show the results from \citet{hilker2017}, who observed the system using FORS2 `onion-shell'-like multi-slit spectroscopy to study the wide range of velocities in the extended stellar halo. To make the comparison of the long-slit data with 2D observations, we used data points whose luminosity-weighted center fall within pseudo-slits aligned with the long-slit observations using widths of \SI{20}{\arcsec} (\SI{5.24}{\kilo\parsec}) and \SI{40}{\arcsec} (\SI{10.48}{\kilo\parsec}) for the MUSE and FORS2 data respectively. This comparison indicates that our analysis is in good agreement with previous observations.

There is considerable scatter in \ac{LOS} velocity and velocity dispersion in Fig. \ref{fig:kinmajaxis} at radii $>10$ kpc in both the southeast and northwest directions and the approach to the cluster values is not smooth. Following previous analysis by \citet{2011A&A...528A..24V} and \citet{2016A&A...589A.139B}, our current understanding of the scatter of the first (velocity) and second (velocity dispersion) moment measurements of the LOSVD is related to the presence of substructures, despite the masking of the localized high(-er) surface brightness sources. As shown in \citet{2012A&A...545A..37A} and \citet{2011A&A...528A..24V}, there is a group of dwarf galaxies with positive LOS velocities ($V_{\rm sys} > 4000$ \si{\kilo\meter\per\second}) that is seen in projection on top of the offset envelope \citep[see figures 14 and 18 in][]{2012A&A...545A..37A}. In addition to the dwarfs, there is the tidal tail off HCC 026 ($V_{\rm sys} = 4946$ \si{\kilo\meter\per\second}) that falls within our MUSE B pointing. The residual unmasked light from these satellites contributes to the averaged light of the different Voronoi bins, and increase the scatter in the measurements of the LOSVD moments.  For example, contribution from these high velocity dwarfs will shift the measured first moments of the LOSVD to higher values, but their intrinsic velocity dispersion is much smaller than that of the diffuse envelope at the same distance for the center. See the results of the spectra decomposition in \citet{2012A&A...545A..37A}, Fig. 16, for further assessment of this point. As our goal is to characterize the large scale variations of the LOSVD moments in the halo and envelope with radius, we do not wish to model each single velocity measurement, i.e. the scatter, but the large scale variations only.

\subsection{Systemic velocity}
\label{sec:starvel}

\begin{figure}
\centering
\includegraphics[width=0.99\linewidth]{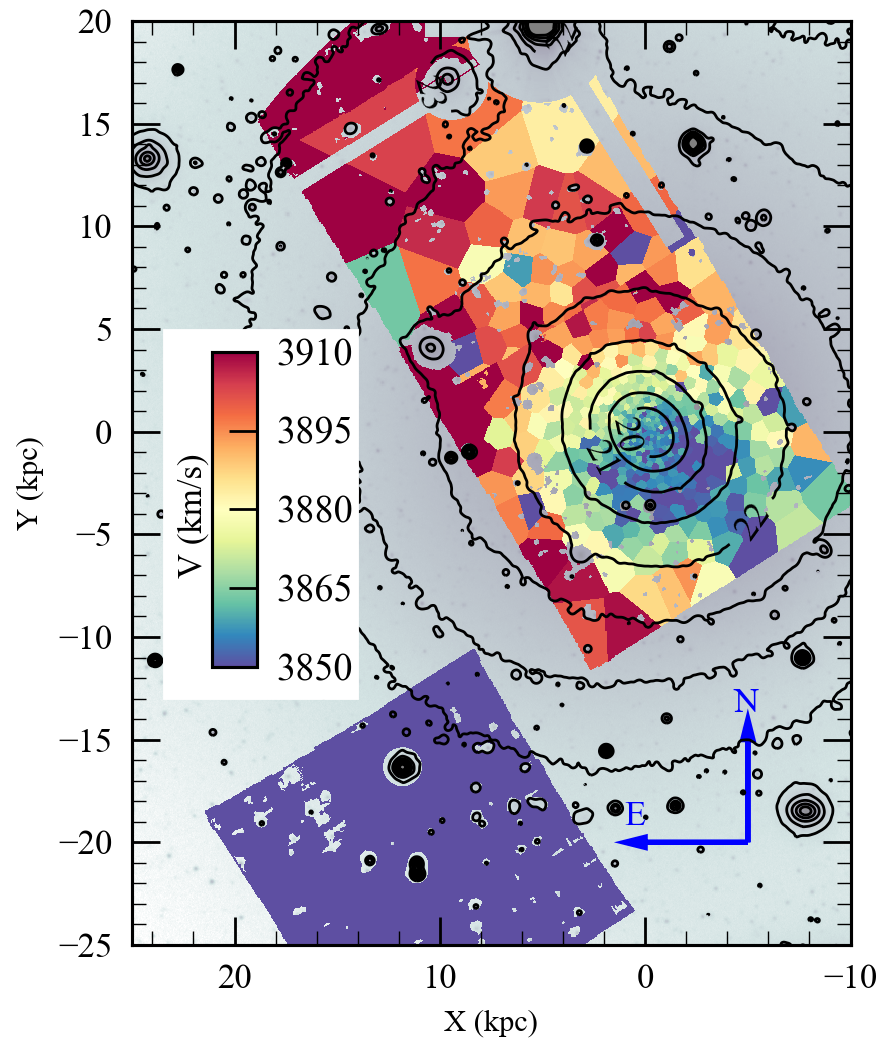}
\caption{Map of the LOS velocity ($V$) for NGC~3311 for all bins with S/N$>10$. The typical observational uncertainty for the LOS velocity is \SI{6}{\kilo\meter\per\second}.}
\label{fig:velmap}
\end{figure}

Fig.~\ref{fig:velmap} shows the systemic velocity of NGC~3311 obtained with our MUSE observations. We have avoided poor fittings (S/N$<10$) and we also do not show the systemic velocity of dwarfs and other compact system identified in our analysis to highlight only the velocity field of NGC~3311. The typical uncertainty of \SI{6}{\kilo\meter\per\second} is calculated as the mean uncertainty considering all the data. 

In Fig.~\ref{fig:velmap}, the first important result is the difference between the systemic velocity in the galaxy center compared to the velocity in the outer regions. In a previous study of NGC~3311 \citep{2016A&A...589A.139B}, we showed that the stellar population properties indicate a separation between the central galaxy, and the cD envelope. A division between these two populations is the V-band surface brightness level of $\mu_V=22$ mag arcsec$^{\text{-2}}$ or, alternatively, the galactocentric distance of $R\approx R_e=8.4$ kpc. We compute the average velocity of the central galaxy within $R<R_e/8$ to be  $3858 \pm 5$ \si{\kilo\meter\per\second}. The average velocity of the cD envelope from \ac{LOS} velocity measurements in the region $R>R_e$ is $V_{\rm cD}=3894\pm 14$ \si{\kilo\meter\per\second}. Based on these results, the relative radial velocity of the galaxy in relation to the outer stellar halo is $\Delta V = 36\pm 15$ \si{\kilo\meter\per\second}, measured solely by the variation of $\sigma$ in the integrated stellar light. We return to this point in Section~\ref{sec:pecvel}, where we derive the peculiar velocity of NGC~3311 based on our combined modeling of the photometric and kinematic data.

\begin{figure}
\centering
\includegraphics[width=0.98\linewidth]{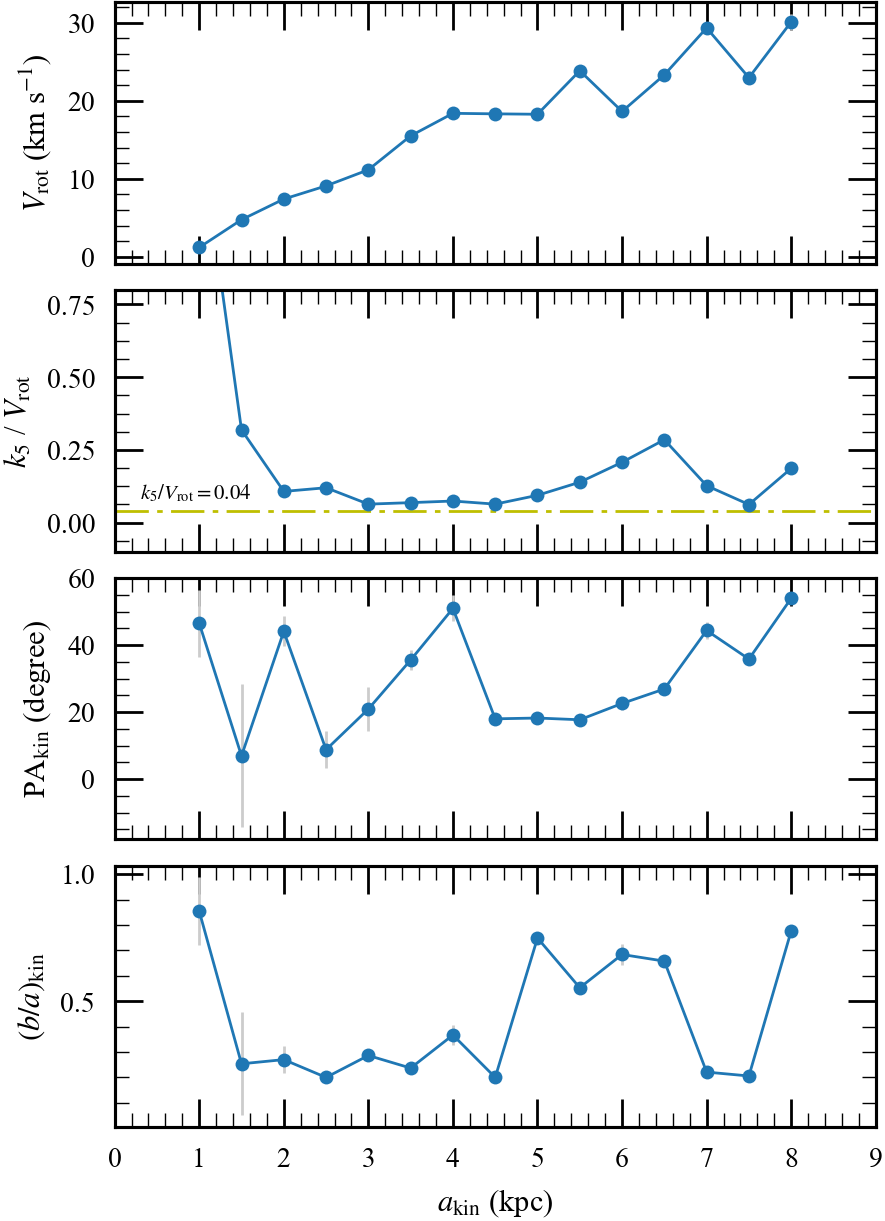}
\caption{Parameters of the rotation model calculated with \textsc{kinemetry} to describe the MUSE velocity field for NGC~3311. From top to bottom, the panels describe the rotation velocity ($V_{\text{rot}}$), the kinematic position angle (PA$_{\text{kin}}$) and minor-to-major axis ($(b/a)_{\text{kin}}$) and the fifth-to-first harmonic weight of the model($k_5/V_{\text{rot}}$) as a function of the kinematic semi major-axis $a_{\rm kin}$. The horizontal dashed line in the bottom panel indicates the maximum threshold value of $0.04$, for early-type galaxies set by the \textsc{ATLAS}$^{\text{3D}}$ survey. If the cosine function is a good approximation for the 2D velocity field, the  $k_5/V_{\text{rot}}$ measured values would lie below this line.}
\label{fig:kinemetry}
\end{figure}

Another important result that can be derived from our 2D velocity mapping is the amount of the rotation of the system and its angular momentum, which are connected to its formation and merging history. In our \ac{MUSE} observations, we are able to study the rotation properties mostly in field I, that contains the central galaxy, but see \citet{hilker2017} for a detailed study these properties at large radius.  We use the program \textsc{kinemetry} \citep{2006MNRAS.366..787K}, that generalizes the methodology of ellipse fitting  with harmonic expansions to obtain surface brightness profiles of galaxies \citep[e.g.,][]{1987MNRAS.226..747J} to the velocity distribution for a rotating system. Fig.~\ref{fig:kinemetry} summarizes the results, indicating the kinematic semi-major axis ($a_{\rm kin}$) profiles for the rotation velocity ($V_{\text{rot}}$), the kinematic position angle (PA$_{\text{kin}}$) and minor-to-major axis ratio ($(b/a)_{\text{kin}}$), and the fifth-to-first harmonic ratio ($k_5/V_{\text{rot}}$). 

The rotation velocity for the central galaxy increases as a function of the radius, but it is only \SI{30}{\kilo\meter\per\second} at its maximum at $R\sim R_e$, which is indeed smaller than the lowest velocity dispersion measured at the galaxy centre ($\sigma_0=175$ \si{\kilo\meter\per\second}). This result together with the small ellipticity of the system  \citep[$\varepsilon=0.05$,][]{2012A&A...545A..37A} indicates an inner galaxy not supported by rotation, according to a simple $V/\sigma$ vs. $\varepsilon$ classification \citep[e.g.,][]{2007MNRAS.379..401E}. Moreover, the rotation obtained in this analysis is not even a good representation of the overall velocity map, as indicated by the ratio between the fifth harmonic term and the rotation velocity ($k_5 / V_{\text{rot}}$). In \textsc{kinemetry}, $k_5$ represents the first term that is not fitted as part of the cosine expansion of the velocity field, and thus indicates the amount of velocity gradient along the LOS that is not included in a rotation model. As a reference, the \textsc{ATLAS}$^{\text{3D}}$ survey, that investigated a volume-limited sample of 260 early-type galaxies \citep{2011MNRAS.414.2923K}, adopted the threshold of $k_5/V_{\text{rot}}=0.04$ as the maximum ratio for which a cosine model is a good description of the projected 2D velocity field. However, in the case of NGC~3311, $k_5 / V_{\text{rot}} > 0.04$ at all radii in the inner galaxy. Therefore, NGC~3311 is a slow rotator, similarly to other galaxies in the same mass class, in agreement with the recent findings from the MASSIVE survey \citep{2017MNRAS.464..356V}.

Besides the kinematic classification of NGC~3311, the \textsc{kinemetry} analysis indicates that substructures may have an important role in the description of the velocity field of this galaxy. This is hinted at by the occurrence of the variations in the kinematic position angle and axis ratio of the rotation model, as function of the radial distance. For instance, in the inner region ($a_{\rm kin}<2$ kpc) there is low support for the rotation model, and the kinematic position angle and axis ratio have strong variations. However, in an intermediate range ($2\lesssim a_{\rm kin}\lesssim 4.5$ kpc), the ratio $k_5/V_{\text{rot}}$ approaches the value expected for a regular rotator, with a constant axis ratio $(b/a)_{\rm kin}\approx 0.25$ and a regular change in the kinematic position angle. Then, in the next following radial range  ($4.5\lesssim a_{\rm kin}\lesssim 7$ kpc), the rotation velocity becomes larger, but the kinematic position angle and axis ratio are approximately constant. Such variations in the geometric parameters of the radial profiles are typically found in surface brightness profiles of spiral galaxies \citep[e.g.,][]{2015MNRAS.453.2965B}, where bulges, disks and other components overlap in the \ac{LOS} to produce the surface brightness profiles.

\subsection{Velocity dispersion}
\label{sec:sigma}
\begin{figure}
\centering
\includegraphics[width=0.99\linewidth]{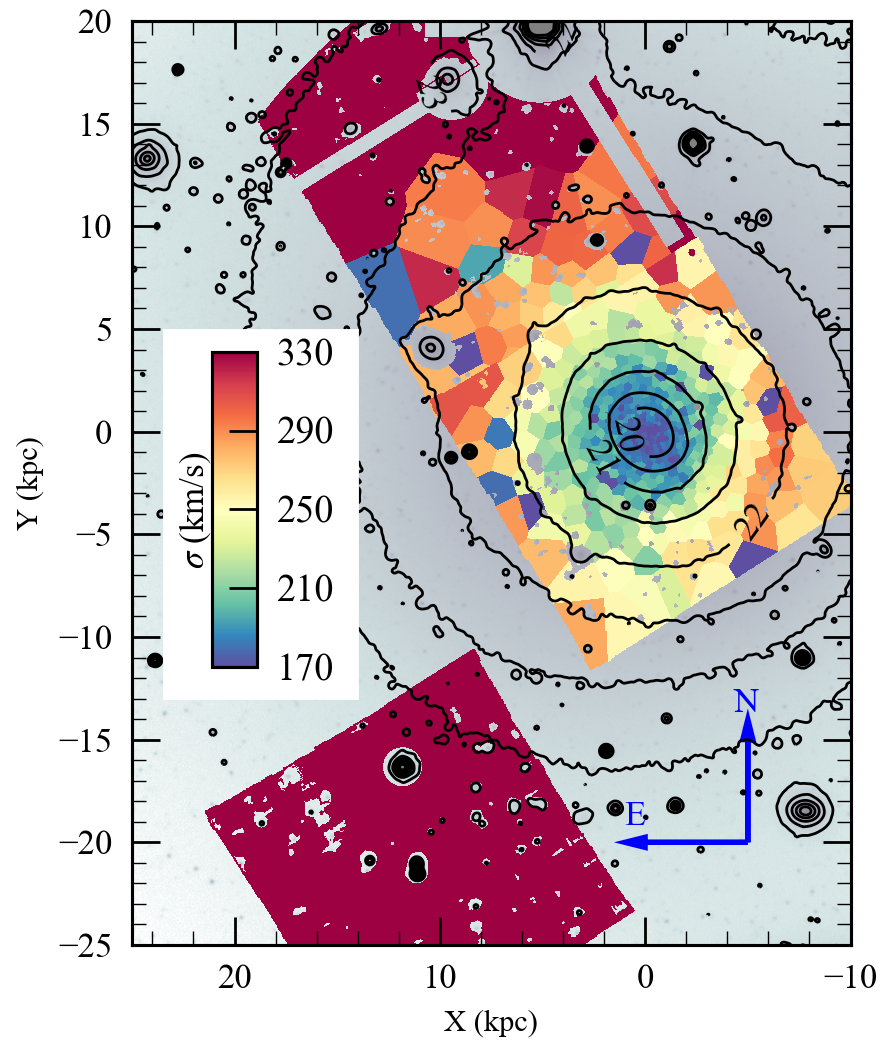}
\caption{Same as Fig.~\ref{fig:velmap} for the LOS velocity dispersion ($\sigma$) of NGC~3311. The typical observational uncertainty for the LOS velocity dispersion is \SI{12}{\kilo\meter\per\second}.}
\label{fig:sigmap}
\end{figure}

Fig.~\ref{fig:sigmap} shows the 2D velocity dispersion ($\sigma$) field of NGC~3311. Typical uncertainties in the measurements of the velocity dispersion are of \SI{12}{\kilo\meter\per\second}. The most important feature of Fig.~\ref{fig:sigmap} is the rising velocity dispersion with radius from the center of the galaxy, which is consistent with long-slit spectroscopy results \citep{2008MNRAS.391.1009L, 2010A&A...520L...9V, 2011A&A...531A.119R}. In the \ac{MUSE} data, we measure a central velocity dispersion of $\approx 175$\si{\kilo\meter\per\second}, reaching values of $\approx 750$\si{\kilo\meter\per\second} in the outermost data points (see also Fig. \ref{fig:kinmajaxis}). 

Positive gradients for the LOS velocity dispersion profiles were reported in several \ac{ETGs} \citep[e.g.,][]{2008MNRAS.391.1009L,2009MNRAS.394.1249C}, including \ac{BCG}s such as IC~1101 \citep{1979ApJ...231..659D}, M87 \citep{2014ApJ...785..143M} and NGC~6166 \citep{2015ApJ...807...56B}. More recently, \citet{2017MNRAS.464..356V,2017arXiv170800870V} showed that such velocity dispersion profiles are common among luminous \ac{ETGs}. However, it is challenging to determine what the maximum velocity dispersion is from these observations and, in particular, whether these velocity dispersions obtained by integrated light reach the cluster's velocity dispersion values. This is directly observed in NGC~3311, as indicated in Fig.~\ref{fig:kinmajaxis}, as both the \ac{MUSE} (this work) and the FORS2 data \citep{hilker2017} reach the cluster's velocity dispersion value of $\sigma=724$ \si{\kilo\meter\per\second} \citep{2003ApJ...591..764C}. 

One interesting property of the 2D velocity dispersion field is the large-scale asymmetry along the photometric major-axis. The second panel of Fig.~\ref{fig:kinmajaxis} shows that the velocity dispersion profile rises faster and reaches the cluster velocity dispersion towards the northeast direction (negative $R$). This result indicates the effect of the large northeast structure observed both in X-ray \citep{2004PASJ...56..743H} and V-band imaging \citep{2012A&A...545A..37A}. However, considering only the most central region, $|R|<10$ kpc, the velocity dispersion profile has a steeper gradient in the opposite direction, that is, towards the southwest direction. When examining the systemic velocity (Fig.~\ref{fig:velmap}) and velocity dispersion maps (Fig.~\ref{fig:sigmap}), close to the isophote contour at $\mu_v = 22$ mag arcsec$^{-2}$ in the southeast most corner of the MUSE pointing I, we identify a subregion where the $\sigma$ values are in the range $255 - 300$ \si{\kilo\meter\per\second} and the LOS velocities are locally blueshifted, with respect to the average values along the same isophote. We consider it as an indication of the presence of kinematic substructure in the \ac{LOSVD} at that location. In section~\ref{sec:photometry} we will argue that this local variation of the velocity dispersion and LOS velocity is related to a new photometric substructure that is found in our analysis for the first time. 

\subsection{Skewness $h_3$ and kurtosis $h_4$ 2D maps}

\begin{figure}
\centering
\includegraphics[width=0.99\linewidth]{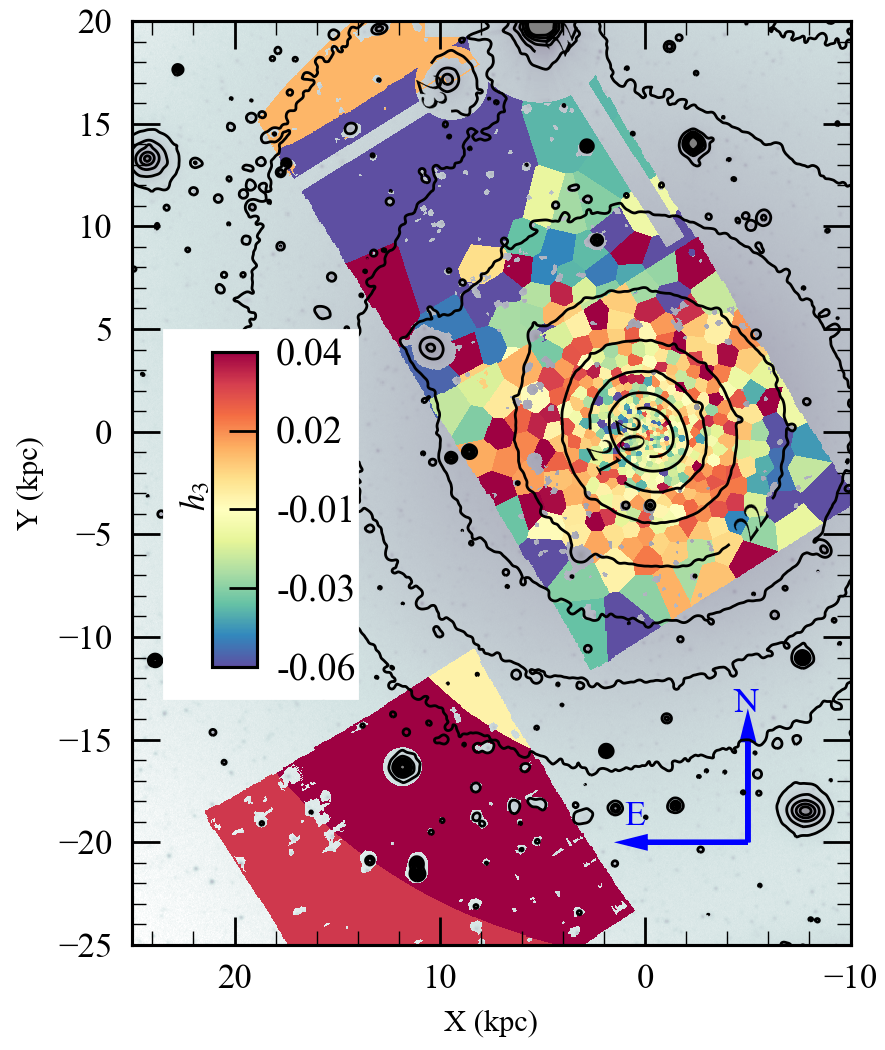}
\caption{Same as Fig.~\ref{fig:velmap} for the skewness parameter ($h_3$) of NGC~3311. The typical observational uncertainty for $h_4$ is $0.02$.}
\label{fig:h3map}
\end{figure}

\begin{figure}
\centering
\includegraphics[width=0.99\linewidth]{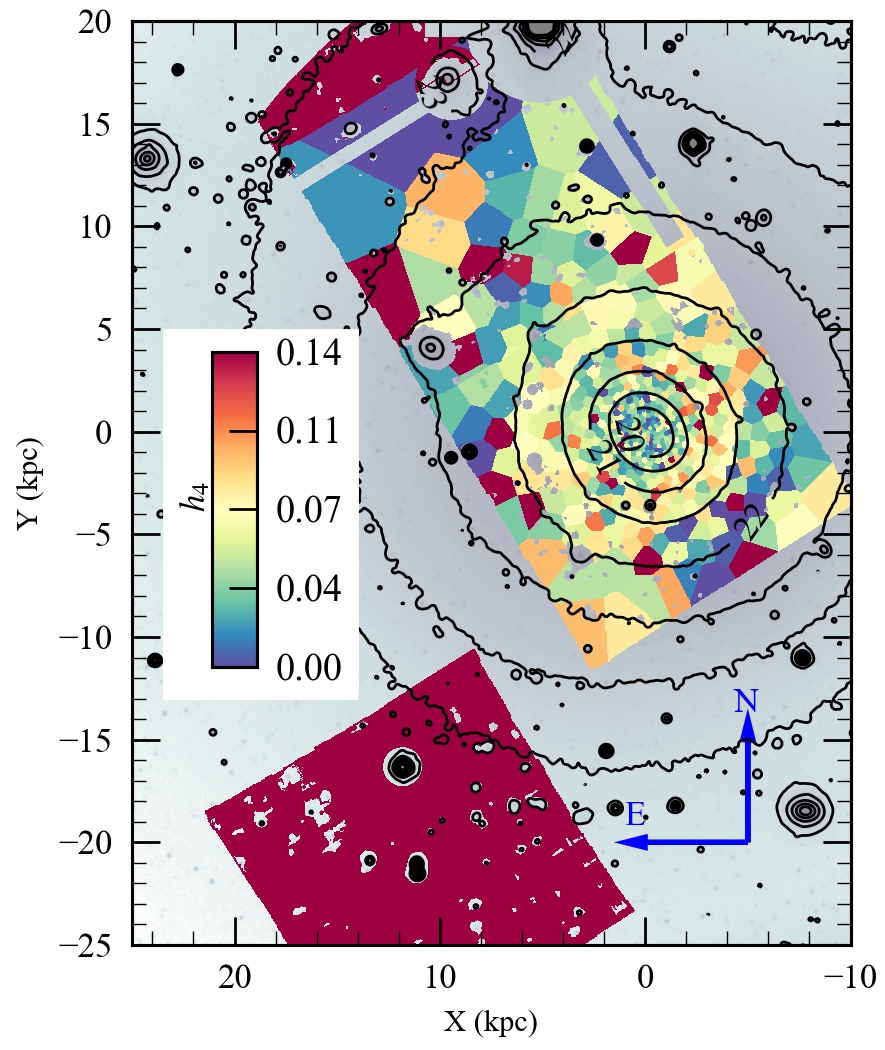}
\caption{Same as Fig.~\ref{fig:velmap} for the kurtosis parameter ($h_4$) of NGC~3311. The typical observational uncertainty for $h_3$ is $0.03$.}
\label{fig:h4map}
\end{figure}

The higher-order deviations from a Gaussian \ac{LOSVD} are measured through the third ($h_3$) and fourth ($h_4$) order coefficients of the Gauss-Hermite expansion of the LOSVD \citep[see][]{1993MNRAS.265..213G,1993ApJ...407..525V}. Figs.~\ref{fig:h3map} and \ref{fig:h4map} show the 2D maps of $h_3$ and $h_4$ from our \ac{MUSE} observations, whose typical uncertainties are 0.02 and 0.03 respectively. The parameter $h_3$ is proportional to the skewness and measures the asymmetric (odd) deviations of the \ac{LOSVD}, i.e., whether the distribution has a blueshifted ($h_3<0$) or redshifted ($h_3>0$) tail. The parameter $h_4$ is proportional to the kurtosis of the \ac{LOSVD} and probes the symmetric (even) deviations of the \ac{LOSVD}, indicating either a more peaked ($h_4>0$) or top-hat ($h_4<0$) shape in relation with a Gaussian distribution. In steeply falling density regions, such as most of the regions covered in the \ac{MUSE} observations with the exception of the center, $h_4$ is also related to radial ($h_4>0$) or tangential ($h_4<0$) anisotropies.

The distribution of the values of the skewness parameter $h_3$ in the area covered by our MUSE pointings has a median value of $-0.01$ with a standard deviation of $0.04$ and there is no (anti)correlation with the velocity gradient of Fig.~\ref{fig:velmap}. However, we note the presence of an annulus with slightly positive values of $h_3$ in the surface brightness range $21\leq\mu_V\leq22$ in Fig.~\ref{fig:h3map} that indicates a steeper leading tail of the \ac{LOSVD} at this location and a variation of $h_3$ with radial distance from the NGC~3311 center. The 2D map of the kurtosis parameter $h_4$ also shows variation with radial distance: the $h_4$ values are near zero in the very central regions, at the northern edge of pointing II and at the southern edge of pointing I, but are otherwise always positive, in the range $0.03 - 0.1$.  The overall distribution of $h_4$ values has a positive median value of $0.05$ and a standard deviation of $0.03$. 

Besides the spatial distribution of $h_3$ and $h_4$, their correlation with $V/\sigma$ contain information about the merging history of \ac{ETGs}. For instance, if the galaxy is a regular rotator, the LOSVDs are asymmetric with the prograde tail being steeper than the retrograde ones, i.e. the degree of asymmetry measured by $h_3$ correlates with $V/\sigma$ \citep{1994MNRAS.269..785B}. However, such correlations are erased in the occurrence of dry major mergers \citep{2014MNRAS.444.3357N}. 

\begin{figure}[t]
\centering
\includegraphics[width=0.99\linewidth]{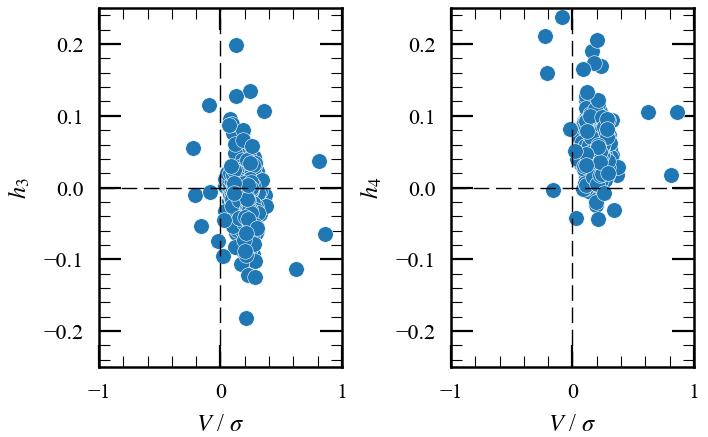}
\caption{Correlation between the Gauss-Hermite high-order moments $h_3$ (left) and $h_4$ (right) with $V/\sigma$.}
\label{fig:h3h4}
\end{figure}

In Fig.~\ref{fig:h3h4} we investigate the correlations of $h_3,\,h_4$ with $V/\sigma$, and found little to none. When comparing our results with simulations from \citet{2014MNRAS.444.3357N}, and considering that we have a round and non-rotating galaxy, we conclude that NGC~3311 resembles those galaxies classified as {\it F type} in \citet{2014MNRAS.444.3357N}, i.e., galaxies formed mostly in dry mergers of already passive galaxies. However, the $h_4$ distribution vs. $V/\sigma$ is remarkably offset: the $h_4$ values scatter around a mean value of 0.05, while they are scattered around zero in simulations. The occurrence of positive $h_4$ values is also observed in other massive \ac{ETGs} \citep{2017MNRAS.464..356V,2017ApJ...835..104V}. In Section~\ref{sec:discussion}, we will discuss this issue further, in the framework of our modeling for the kinematics of NGC~3311.


\section{Photometric model for the surface brightness distribution in NGC~3311}
\label{sec:photometry}

So far, we focused primarily on the analysis of the kinematics of NGC~3311. As a next step, we wish to combine the above results with those from the imaging data to study the late mass assembly of the cD galaxy NGC~3311. For this purpose, we revisit the work of \citet{2012A&A...545A..37A}, who performed a 2D-modeling of the surface brightness in the core region of the Hydra~I cluster using deep V-band photometry, that led to the identification of a large, diffuse, off-centered envelope around NGC~3311.  In this work, we extend the analysis to include the central region of the galaxy also to obtain a more accurate photometric decomposition for NGC~3311 light.
%
%

\subsection{Data and methods}

We use Johnson V-band imaging acquired with FORS1 \citep{1998Msngr..94....1A} at the \ac{VLT} for the observing programme 65.N-0459(A) (PI: Hilker), that we retrieve via the ESO Science Archive Facility. This image covers a field-of-view of $6.8' \times 6.8'$ slightly off-centered from NGC~3311, in order to avoid a bright star towards the northeast direction. We adopted the zero point ZP$=27.43\pm0.06$ from \citet{2012A&A...545A..37A}, and the extinction correction of $0.25$ mag derived from \citet{1998ApJ...500..525S}.

We used the \textsc{sextractor} program \citep{1996A&AS..117..393B} to detect sources in the field-of-view and produce a mask for the photometric fitting.  Bright stars and several dwarf galaxies projected onto the main galaxies were masked manually. We also masked the dust lane at the center of NGC~3311. We used bright, not saturated stars in the field-of-view to estimate the \ac{PSF}, that was calculated  using the \textsc{IRAF} task \textsc{psfmeasure} \citep{1986SPIE..627..733T}, adopting a \citet{1969A&A.....3..455M} profile. The resulting \ac{PSF} has a mean power-law slope of $\beta=2.956\pm0.09$ and a \ac{FWHM} of $0''.87\pm0''.09$. 

We used the \textsc{galfitm} software \citep{2013MNRAS.435..623V}, an updated version of \textsc{galfit} \citep{2002AJ....124..266P}, to perform a parametric structural decomposition of the three main galaxies in the core of Hydra~I, including the cD galaxy NGC~3311, the large elliptical NGC~3309 situated Northwest of NGC~3311 in Fig.~\ref{fig:strategy}, and the S0 galaxy HCC~007 to the South of NGC~3311. We provide further details of the fitting method in what follows.

We built the \textsc{galfitm} models from the single S\'ersic profiles of the three galaxies, and include additional structural components, described with a S\'ersic parametric law, as required from the inspection of the results and the residuals of the model. This iterative method is similar to that applied by \citet{2013ApJ...766...47H}, who showed definite improvements in the decomposition of \ac{ETGs} surface brightness when additional components are added to single S\'ersic models. A priori, we had no particular constraints on the number of components, as our intent is to gain a good overall description of the light of NGC~3311, and we adopt the S\'ersic law because of its flexibility. We judged the photometric models on the basis of the optimization of the $\chi^2$ between models with the goal to achieve the maximum symmetric contribution to the light in NGC~3311 and identify any non-symmetric features.

Each S\'ersic component has seven free parameters, including four to describe the geometry -- central coordinates $X$ and $Y$, position angle of the semi-major axis $PA$ and minor-to-major axis ratio $q=b/a$ -- which are then used to calculate the isophotal distance $R$, plus three parameters to describe the light profile given by \citep{1968adga.book.....S}

\begin{equation} 
I(R)=I_e\exp \left \{ -b_n \left [ \left ( \frac{R}{R_e}\right )^{1/n}-1\right ]\right \} \mbox{,}
\label{eq:sersic} 
\end{equation}

\noindent where $R_e$ is the effective radius, $I_e$ is the intensity at the effective radius, $n$ is the S\'ersic index and $b_n\approx 2n - 0.32$ is a parameter that ensures that $R_e$ contains half the light of the galaxy \citep[see][]{1999A&A...352..447C,2003ApJ...582..689M}.  

The sky subtraction is an important source of error in the photometric decomposition \citep[see][]{1996A&AS..118..557D}, and the V-band imaging does not contain any obvious regions where the sky brightness can be measured independently, as most of the field-of-view contains light from either the cD halo or possible other faint \ac{ICL} structures, such as streams and tidal tails \citep[see ][]{2005ApJ...631L..41M,2017ApJ...839...21I}. Therefore, we performed the sky subtraction along with the photometric modeling with \textsc{galfitm} by including a constant sky component as a free parameter to the fitting. Simulations by \citet{2007ApJS..172..615H} showed that the use of a variable sky level does not affect the parameters of the S\'ersic profiles strongly.

The main advantage of \textsc{galfitm} is the inclusion of a non-parametric component in the fitting, that is suitable to unveil faint and non-axisymmetric structures, such as tidal streams, by finding models for which the residuals are mostly positive. Interestingly, this idea is similar to the maximum symmetric model from \citet{2012A&A...545A..37A}, that allowed the identification of the off-centered envelope around NGC~3311. 

\subsection{Photometric components for the NGC~3311 extended light distribution}
\label{sec:sercomponents}

\begin{figure*}[t]
\centering
\includegraphics[width=0.98\linewidth]{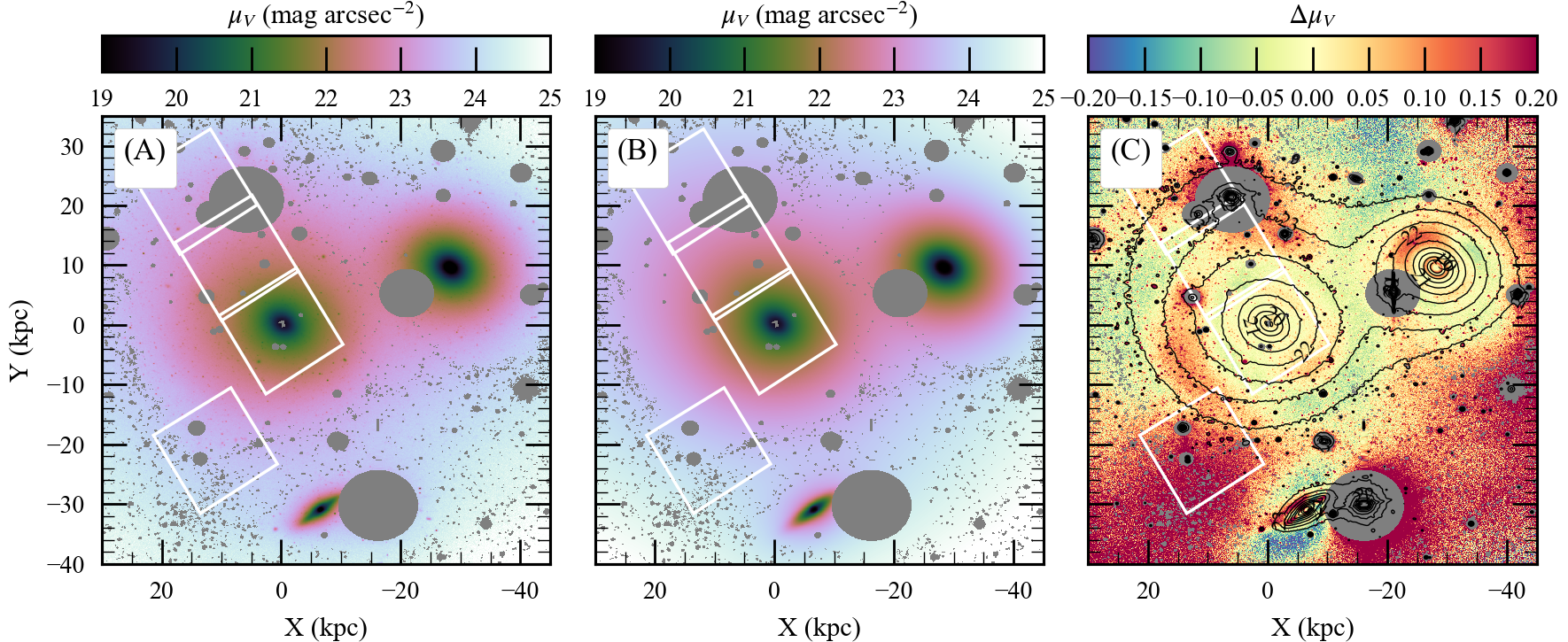}
\caption{Photometric model of the three main galaxies in the Hydra~I cluster core in FORS1 V-band image. Left: Observed V-band image calibrated with the zero-point from \citet{2012A&A...545A..37A}. Center: \textsc{galfitm} model including the ten S\'ersic components listed in Table~\ref{tab:galfitpars}, obtained with the non-parametric fitting method. Right: Residual image (original minus model), normalized by the sigma image which contains the error and noise of the observations.}
\label{fig:photV}
\end{figure*}

Our final \textsc{galfitm} model for the light in the V band image of NGC~3311 is shown in Fig.~\ref{fig:photV}. It contains ten S\'ersic components, whose parameters are listed in Table~\ref{tab:galfitpars}. HCC~007 and NGC~3309 required three components each, and NGC~3311 required four components. We notice that the residual map has several positive, rather than negative, features, that comes as the result of the non-parametric components from \textsc{galfitm}. Considering that we have build a model based solely on the quality of the fitting, the model components are not necessarily associate to distinct physical structures. However, the number of components for each galaxy in our model is in agreement with results from the literature, as discussed below. Moreover, in Section \ref{sec:fmdapplication}, we will show that the kinematics of these systems support the idea that some of these structures are actually physically motivated.

\begin{table*}
\caption{Galfit parameters for the photometric model using S\'ersic components. (1) Identification of the galaxy. (2) Identification of the photometric component. (3-4) Central X and Y coordinates of the S\' ersic profiles relative to the center of the galaxies: NGC~3311 at $\alpha (J2000) $=10h36m42.7s and $\delta (J2000) =$-27d31m40.8s; NGC~3309 at $\alpha (J2000) =$10h36m35.6s and $\delta (J2000) =$-27d31m04.91s; and HCC~007 at $\alpha (J2000) =$10h36m41.18s and $\delta (J2000) =$-27d33m39.2s. (5) Absolute magnitude. (6) Effective radius. (7) S\'ersic index. (8) Minor-to-major axis ratio. (9) Position angle of the major axis. (10) Fraction of the light of the component relative to all other components in the same galaxy.}             
\label{tab:galfitpars}   
\centering
\scriptsize
\begin{tabular}{cccccccccc}
\hline
\hline
Galaxy & ID & X (kpc) & Y (kpc) & $M_{\rm tot}$ & $R_e$ (kpc) &  n & $b/a$ & PA (degree) & $f$\\ 
(1) & (2) & (3) & (4) & (5) & (6) & (7)  & (8) & (9) & (10)\\
\hline
\multirow{4}{6em}{NGC 3311}& A & $0.00 \pm 0.04$ & $0.00 \pm 0.07$ & $-18.2 \pm 0.5$ & $1.40 \pm 0.06$ & $0.51 \pm 0.05$ & $0.790 \pm 0.018$ & $54 \pm 7$ & $0.008$\\
  & B & $0.14 \pm 0.05$ & $0.31 \pm 0.05$ & $-19.56 \pm 0.20$ & $2.96 \pm 0.17$ & $0.69 \pm 0.16$ & $0.820 \pm 0.012$ & $36.9 \pm 2.3$ & $0.030$\\
  & C & $-0.33 \pm 0.06$ & $-0.58 \pm 0.15$ & $-21.29 \pm 0.10$ & $9.60 \pm 0.31$ & $0.85 \pm 0.04$ & $0.970 \pm 0.012$ & $-55 \pm 14$ & $0.147$\\
  & D & $4.4 \pm 0.4$ & $7.4 \pm 0.5$ & $-23.15 \pm 0.12$ & $51 \pm 5$ & $1.04 \pm 0.07$ & $0.890 \pm 0.012$ & $59.7 \pm 3.5$ & $0.815$\\
\hline
\multirow{3}{6em}{NGC 3309}& E & $0.000 \pm 0.006$ & $0.000 \pm 0.006$ & $-18.44 \pm 0.06$ & $1.174 \pm 0.010$ & $0.150 \pm 0.012$ & $0.830 \pm 0.006$ & $47.8 \pm 1.8$ & $0.052$\\
  & F & $0.057 \pm 0.016$ & $0.094 \pm 0.015$ & $-19.85 \pm 0.10$ & $2.451 \pm 0.034$ & $0.540 \pm 0.030$ & $0.820 \pm 0.006$ & $43.7 \pm 1.2$ & $0.190$\\
  & G & $-0.07 \pm 0.04$ & $-0.22 \pm 0.04$ & $-21.345 \pm 0.024$ & $7.96 \pm 0.15$ & $1.13 \pm 0.05$ & $0.920 \pm 0.006$ & $54.2 \pm 3.2$ & $0.758$\\
\hline
\multirow{3}{6em}{HCC 007}& H & $0.000 \pm 0.010$ & $0.000 \pm 0.007$ & $-16.76 \pm 0.04$ & $0.539 \pm 0.012$ & $0.230 \pm 0.018$ & $0.800 \pm 0.024$ & $-32 \pm 4$ & $0.050$\\
  & I & $0.007 \pm 0.023$ & $-0.001 \pm 0.017$ & $-18.60 \pm 0.04$ & $2.41 \pm 0.07$ & $0.830 \pm 0.030$ & $0.310 \pm 0.006$ & $-51.0 \pm 0.5$ & $0.274$\\
  & J & $-0.50 \pm 0.06$ & $-0.044 \pm 0.035$ & $-19.6 \pm 0.4$ & $15 \pm 8$ & $3.3 \pm 0.6$ & $0.688 \pm 0.030$ & $-51 \pm 7$ & $0.676$\\
\hline
 \hline
\end{tabular}
\normalsize
\end{table*}

{\it Three S\'ersic components for \ac{ETGs} $-$}
The components that are required to match the 2D light distribution in our model are consistent with those discussed in \citet{2013ApJ...766...47H} for a sample of isolated ETGs. These authors found that ETGs may be decomposed into three, and up to four, structural components when a bi-dimensional parametric fitting is carried out to reproduce the semi-major axis position angle and the ellipticity profiles consistently, in addition to the surface brightness profile. Our model returns values for the effective radii that are in good agreement with those from \citet{2013ApJ...766...47H}, whose typical values are $R_e\lesssim 1$ \si{\kilo\parsec}, $R_e\approx 2.5$  \si{\kilo\parsec}  and $R_e\approx 10$  \si{\kilo\parsec} for decomposition with three components. 

\begin{figure}[t]
\centering
\includegraphics[width=\linewidth]{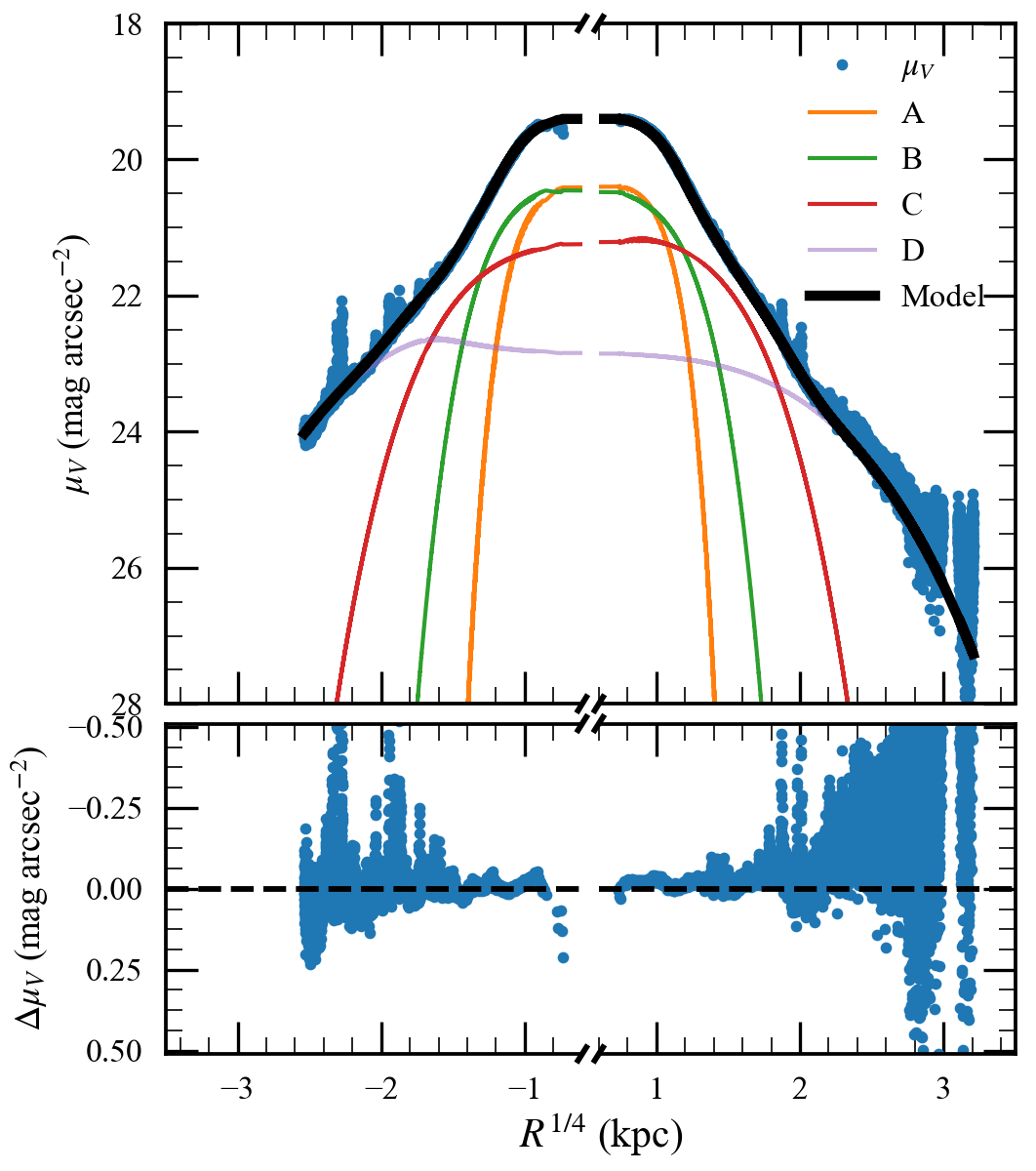}
\caption{Surface brightness profile along the photometric major axis at $P.A.=$\SI{55}{\degree}. \textit{Top:} Comparison of the V-band surface brightness (blue dots) with the best fit from \textsc{galfitm} (black solid line). Other solid lines indicate the surface brightness of the subcomponents indicated in Table \ref{tab:galfitpars} including A (orange), B (green), C (red) and D (purple). \textit{Bottom:}  Residuals between the V-band surface brightness and the model. Positive (negative) radius indicate the distance from the center of NGC 3311 towards the southeast (northwest) direction.}
\label{fig:sbprofile}
\end{figure}

{\it Extended enevelope around NGC~3311 $-$}
The modeling of the light of NGC~3311 clearly requires an additional component, labeled as ``D'' in Table~\ref{tab:galfitpars}. Component D has a central offset of \SI{8.6}{\kilo\parsec} to the Northeast direction and has a much larger effective radius of \SI{51}{\kilo\parsec} with respect to the other three components. Fig.~\ref{fig:sbprofile} shows the surface brightness profile of NGC~3311 at a fixed position angle of \SI{55}{\degree}, which is approximately aligned with the semi-major axis from photometric component ``A''. This surface brightness profile shows that the off-centered D component is required to explain the excess of light at large galactocentric distances to the North-East of the galaxy center. 

This component, the fourth, matches the off-centered envelope found by \citet{2012A&A...545A..37A}, that is also associated with an X-rays secondary peak \citep{2004PASJ...56..743H,2006PASJ...58..695H} and an excess of metallicity in the stellar populations \citep{2016A&A...589A.139B}. Hence on the basis of the photometry, there is a structural component in the light of NGC~3311 that can be identified with a cD halo. 

{\it Intracluster light in the Hydra~I core and asymmetric substructures in the light of NGC~3311 $-$}
A large scale characterization of the intracluster light (\ac{ICL}) in the Hydra~I cluster would require a larger field-of-view than the one available with the FORS1 V-band image. Nonetheless we would like to describe the three most noticeable features in our V-band image, that are related to the ICL in the core of the cluster, and may have implication for our kinematic analysis. 

\begin{figure}[t]
\centering
\includegraphics[width=\linewidth]{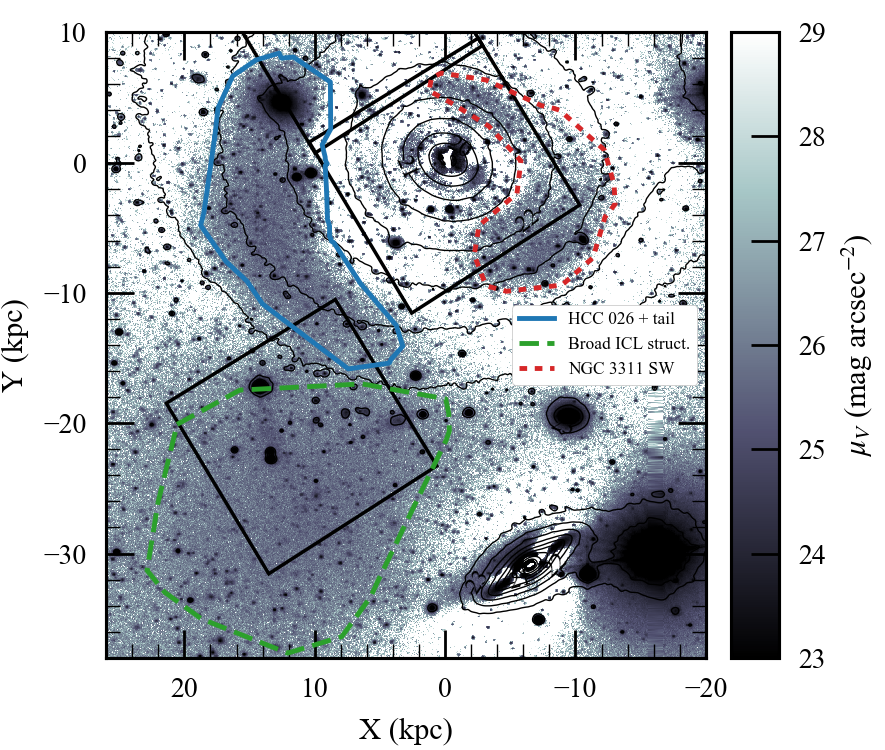}
\caption{\ac{ICL} substructures observed in the residual map from the modeling of symmetric structures with \textsc{galfitm}. V-band contours are also shown for reference. Black squares represent the MUSE fields, and the colored polygons represent the three most evident substructures in this region.}
\label{fig:photres}
\end{figure}

Fig.~\ref{fig:photres} shows a subsection of the residuals that highlights the presence of several \ac{ICL} substructures. The \ac{ICL} is formed by various mechanisms that unbound stars from their parent halos and may contain from 5\% to 50\% of all the stars in clusters \citep[see][]{2010HiA....15...97A,2011ARep...55..383T}. Simulations show the variety of shapes of these structures, including tails, plumes, and shells \citep[see][]{2006ApJ...648..936R,2007MNRAS.377....2M,2009ApJ...699.1518R,2015MNRAS.451.2703C} that have been unveiled observationally with deep photometry \citep[e.g.,][]{2005ApJ...631L..41M,2017ApJ...834...16M,2017ApJ...839...21I}.

A distinct feature is a large tidal tail first discovered by \citet{2012A&A...545A..37A} that is associated with the galaxy HCC~026, a spectroscopically confirmed member of the Hydra I cluster core \citep{2008A&A...486..697M}. In Fig.~\ref{fig:photres}, HCC~026 is in the upper part of the structure marked by the solid blue line; unfortunately, most of its light falls outside the mosaiced area of our \ac{MUSE} observations.     

The large structure indicated by the dashed green line in Fig.~\ref{fig:photres} was also identified by \citet{2012A&A...545A..37A}. They suggested that it originates from a broad tidal tail off HCC~007. Our current study indicates that this structure is broader than previously reported, and it is connected to HCC~007 as indicated by the asymmetry of the residuals to the East of the center of HCC~007, after the subtraction of the parametric model. 

Finally, the residual map displays a substructure located to the Southwest of NGC~3311, marked in Fig.~\ref{fig:photres} by the red dotted line. This substructure is located at the relative position $(-7, 14)$ kpc from the center of NGC~3311, and it falls partially within the field of view of Field~I in our MUSE observations. This structure to the southwest of NGC~3311 may be responsible for the blueshifted LOS velocity and the locally high velocity dispersion observed in the 2D maps in Figs.~\ref{fig:velmap} and \ref{fig:sigmap}. In Section~\ref{sec:modeling}, we provide additional kinematics and photometric evidence for this substructure. The current photometric model does not require a common shared halo including NGC 3311 and NGC 3309. The current \textsc{galfitm} models do not  provide evidence for residual light West of NGC 3309. Constraints for additional light at larger radii may come for future wide field photometric data. 

We conclude this section by stating that the S\'ersic parametric multi-component model produces an overall good match to the surface brightness distribution in the V-band image. The outcome of the current analysis shows the requirement for a fourth component to model the light of NGC~3311. This component has a central offset from the other three, and a very large effective radius (51 kpc), matching the morphological definition of a cD halo for \ac{BCG}s \citep{1964ApJ...140...35M,1965ApJ...142.1364M}. However, a point raised by \citet{2015ApJ...807...56B} questions whether one is able to identify \textit{physically} distinct galaxy components from photometry alone. In the next section, we address this problem in NGC~3311, by coupling the photometry with the 2D kinematics from the new MUSE datacubes.


\section{Photometric and kinematic modeling of NGC~3311}
\label{sec:modeling}

The goal of this section is to construct a model that can combine the 2D \ac{LOSVD}s of NGC~3311 (Section \ref{sec:kinematics}) on the basis of the photometric components matching the V-band surface brightness (Section~\ref{sec:photometry}). Our approach is motivated by the work of \citet{2015ApJ...807...56B}, which we review in the following, before explaining our generalization.

In their work, \citet{2015ApJ...807...56B} studied NGC~6166, the central galaxy of Abell 2199 and another prototypical cD galaxy. They obtained their kinematics from long-slit observations extending out to $100''$ ($\sim 60$ kpc) from the center of the galaxy using deep observations from the Hobby-Eberly telescope. The velocity dispersion profile of NGC~6166 increases with radius, reaching the cluster velocity dispersion value of \SI{819}{\kilo\meter\per\second}, similarly to the velocity dispersion profile NGC~3311. Besides, they revisited the photometric data of the system from a variety of sources to produce an accurate surface brightness profile, which they modeled using decompositions into either one or two S\'ersic components. 

To combine the photometric and kinematic information, \citet{2015ApJ...807...56B} modeled the \ac{LOSVD} profile as a luminosity-weighted superposition of two Gaussian \ac{LOSVD}s, each with its own velocity dispersion, whose luminosity follows the surface brightness of the S\'ersic components. By fixing the velocity dispersion of the combined model to values of $\sigma_{in}=300$ \si{\kilo\meter\per\second} and $\sigma_{out}=865$ \si{\kilo\meter\per\second} for the inner and outer components respectively, they have shown that the best model requires two components with the least overlap in radius. Specifically, the inner component had an effective radius $R_{e,1}=14.95''\approx 9$ kpc and a S\'ersic index $n_1=1.52$, whereas the outer component had $R_{e,2}=181.2''\approx 110$ kpc and $n_2=1.83$. However, the best combined model is not the best model according to the photometry alone, at least from a $\chi^2$ test. \citet{2015ApJ...807...56B} argued that the surface brightness decomposition of the best combined photometric and kinematic model is slightly inconsistent with their photometric data, and thus caution about the use of photometry alone to infer the properties of the stellar halos.

Regarding our case, the photometric parameters of the best combined model from \citet{2015ApJ...807...56B} resemble the parameters of our surface brightness decomposition for NGC~3311 (Table \ref{tab:galfitpars}), i.e. spheroids with low S\'ersic indexes ($n\lesssim 2$) and very different radii for the inner and outer components. Moreover, our photometric decomposition naturally indicates that the envelope surface brightness is approximately exponential, in agreement with previous findings \citep[e.g.][]{2007MNRAS.378.1575S,2011ApJS..195...15D}. Thus, we now ask the following  question: if we suppose that the photometric decomposition of NGC~3311 is identifying structural components, can their kinematics be modeled to reproduce the MUSE LOSVD maps?

To answer this question, we generalize the model used by \citet{2015ApJ...807...56B} in the following ways: 1) we expand the analysis to a 2D approach, which is best suitable to explore the asymmetric surface brightness of NGC~3311, and exploit the capability of the \ac{MUSE} IFU data; 2) we allow non-Gaussian \ac{LOSVD}s for the components using a mixture model to predict the higher-order moments $h_3$ and $h_4$ also; and 3) we consider more than two components to account for all the structures observed in the surface brightness distrbution. In the next sections, we describe our modeling procedure in detail.

\subsection{The finite mixture distribution model}

In this section, we briefly review the mathematical background of a \ac{FMD}, a probabilistic model which represents a probability distribution as a superposition of components, following the description of the model from \citet{9780387329093}, and applied to the study of \ac{LOSVD}s. In this approach, the observed \ac{LOSVD} $\mathcal{L}(v)$ at a given location $\mathbf{X}=(X,Y)$ in the plane of the sky is represented as 

\begin{equation}
\mathcal{L}(v) = \sum_{i=1}^{N}w_i\mathcal{L}_i(v)\mbox{,}
\label{eq:fmd}
\end{equation}

\noindent where $v$ is the \ac{LOS} velocity, $\mathcal{L}_i(v)$ indicate the probability distributions for the $i=1,...,N$ components, and $w_i$ are the weights, which follow the constraints $w_1+...+w_n=1$ and $w_i\geq 0$. In this approach, the quantities that describe $\mathcal{L}(v)$ are related to the expected value ${\rm E}[g(v)]$ of a generic function $g(v)$, defined as

\begin{equation}
{\rm E}[g(v)]=\int_{-\infty}^{+\infty}g(v)\mathcal{L}(v)dv\mbox{.}
\label{eq:ev}
\end{equation}

\noindent Given that the components of the mixture distribution are also probability distributions, we can also define the expected values of the components ${\rm E}_i[g(v)]$ substituting $\mathcal{L}(v)$ with $\mathcal{L}_i(v)$ in the equation above. Therefore, from equations \eqref{eq:fmd} and \eqref{eq:ev}, it follows that 

\begin{equation}
{\rm E}[g(v)] = \sum_{i=1}^N w_i {\rm E}_i[g(v)]\mbox{.}
\end{equation}

The mean \ac{LOS} systemic velocity $\mu$ is obtained with $g(v)=v$, thus

\begin{equation}
\mu = {\rm E}[v] = \sum_{i=1}^N w_i \mu_i\mbox{,}
\end{equation}

\noindent where $\mu_i={\rm E}_i[v]$ are the mean systemic velocities of the components. Similarly, the \ac{LOS} velocity dispersion $\sigma$ is obtained with $g(v)=(v - \mu)^2$, thus

\begin{equation}
\sigma^2= {\rm Var}[v] = \sum_{i=1}^{N} w_i(\mu_i^2 + \sigma_i^2) - \mu^2\mbox{,}
\end{equation}

\noindent where $\sigma_i$ are the velocity dispersions of the components. In the mixture model, the $m$-th central moment is given by 

\begin{equation}
{\rm E}[(v - \mu)^m] = \sum_{i=1}^{N}\sum_{j=0}^{m}\binom{m}{j}w_i(\mu_i-\mu)^{m-j}{\rm E}[(v-\mu_i)^j]\mbox{.}
\label{eq:centralmoments}
\end{equation}

Finally, the skewness $\gamma_3$ and the kurtosis $\gamma_4$ are then defined as 

\begin{equation}
\gamma_3=\frac{{\rm E}[(v-\mu)^3]}{\sigma^3}
\end{equation}

\begin{equation}
\gamma_4=\frac{{\rm E}[(v-\mu)^4]}{\sigma^4}\mbox{.}
\end{equation}

The \ac{FMD} model described so far section does not depend on the parametrization of the \ac{LOSVD}. However, we are interested in using a parametrization that is related to the observation, i.e., Gauss-Hermite distributions, where the parameters $h_3$ and $h_4$ are used to describe the skewness and the kurtosis of the \ac{LOSVD}, instead of $\gamma_3$ and $\gamma_4$. The relation between these quantities are then given by \citep[see the documentation of the task \textsc{xgauprof} for the GIPSY package, ][]{2001ASPC..238..358V}

\begin{equation}
h_3\approx \frac{\gamma_3}{4\sqrt{3}}\mbox{,}
\end{equation}

\begin{equation}
h_4\approx \frac{\gamma_4 - 3}{8\sqrt{6}}\mbox{.}
\end{equation}

\subsection{Application of FMD model to the LOSVDs of NGC~3311}

To apply the \ac{FMD} model, we assume that the observed \ac{LOSVD} at any given position $\mathbf{X}=(X,Y)$ is given by the luminosity-weighted mixture of the components in the surface brightness decomposition, each with its own \ac{LOSVD}. Therefore, the weights are calculated as 

\begin{equation}
w_i(\mathbf{X}) = \frac{I_{V,i}(\mathbf{X})}{\sum_{j=1}^NI_{V,j}(\mathbf{X})}\mbox{,}
\end{equation}

where $I_{V,i}(\mathbf{X})$ is the V-band surface brightness of the component $i=1,...,N$ at $\mathbf{X}$. Implicitly, this simple modeling considers that the light traces the mass with a mass-to-light ratio that is constant within a given component and that this mass-to-light ratio is the same for all components.

\begin{figure}
\centering
\includegraphics[width=\linewidth]{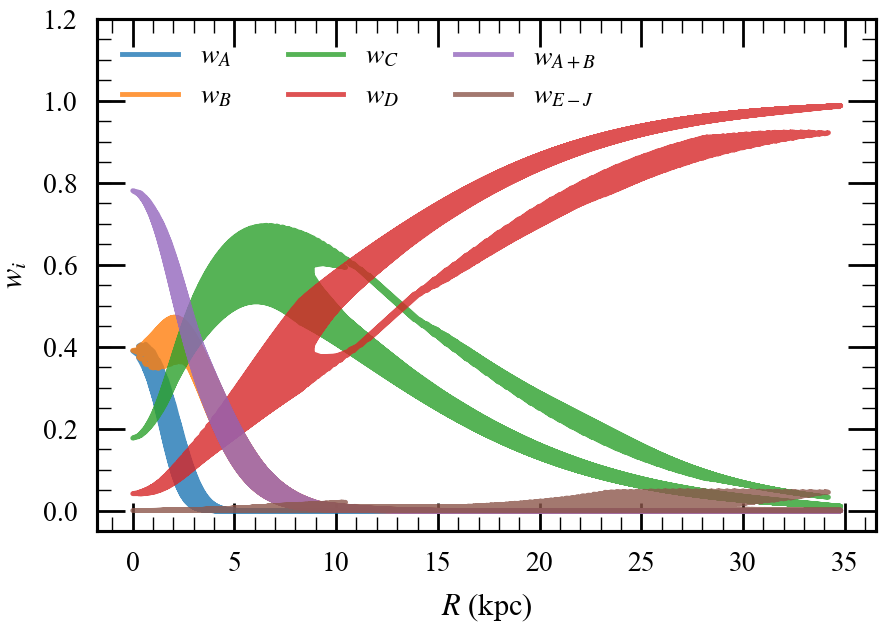}
\caption{Weights for the FMD model of NGC~3311 as a function of the radius within the MUSE fields. The four main components of NGC 3311, A, B, C and D, are shown individually. The component A+B indicates the sum of the first two components which are modeled as having the same kinematics. The component E-J indicates the cumulative contribution of all other photometric components which are not included in the kinematic modeling.}
\label{fig:lightfraction}
\end{figure}

According to the decomposition model described in Section~\ref{sec:photometry}, the surface brightness within the \ac{MUSE} observations is dominated by the four components of NGC 3311. This is illustrated in Fig.~\ref{fig:lightfraction}, which shows the weights of the individual components A to D as a function of the radius for a dense grid of points within fields I to IV. We also show the sum of all modeled components E to J, $w_{E-J}$, which indicates that the sum of the light in other spheroidal components in our observations is not greater than 5\%. We also provide the contribution of the sum of components A and B, indicated as $w_{A+B}$, which are possibly physically associated, as we discuss in the next section. In the actual modeling of NGC 3311, we re-normalize the weights to consider only components A to D.  

%

Next we assume that each component has a unique \ac{LOSVD} characterized by a set of four parameters -- $\mu_i$, $\sigma_i$, $h_{3,i}$, $h_{4,i}$, that do not depend on position on the sky. Consequently, the spatial variation of the modeled \ac{LOSVD} is given only by the linear combination of the weights of the photometric components at the different sky positions. In this way, we are ignoring both ordered motions (rotation) and  bulk motions within a given component. This approach resembles the use of isothermal spheres to describe the \ac{LOSVD}s, where the velocity dispersion does not vary spatially \citep[see][]{1987gady.book.....B}, and we allow these spheres to have non-zero skewness and kurtosis values. 

\subsection{Fitting method}
\label{sec:fmdapplication}

Within the above assumptions, we then calculate $\mu_i$, $\sigma_i$, $h_{3,i}$, and $h_{4,i}$ for components $i=A,...,D$. As indicated in equation~\eqref{eq:centralmoments}, each central moment depends on the previous moments, so we carry out {\it iterative} fittings to calculate each of the set of parameters sequentially. For example, we compute $\mu_i$ directly from the maps, and then we calculated the velocity dispersions ($\sigma_i$) fixing $\mu_i$, and, so on up to $h_{4,i}$. These calculations were performed with the program \textsc{least\_squares} from the SciPy package \citep{Oliphant2007,Perez2011}, using the trust-region reflective gradient method \citep{Branch1999}, which allows a bound-constrained minimization of the parameters, and uncertainties were calculated using bootstrapping simulations. 

We initially computed the \ac{FMD} models using four kinematic components, one for each photometric structure. This approach provided values for $\mu_i$, $\sigma_i$ and $h_{3,i}$ almost identical to the best model that we discuss in the remaining text, but we encountered problems in obtaining a meaningful value of $h_{4,B}$. To improve the fitting, we made two modifications to the procedure. Firstly, we used a robust fitting process instead of a simple weighted $\chi^2$ minimization, i.e. we used an approach that is less sensitive outliers. This was carried out using the minimization of a loss function $\rho(\chi^2)$ instead of the simple minimization of the $\chi^2$. In particular, we adopted a smooth approximation to absolute value implemented in the program \textsc{least\_squares} given by $\rho(\chi^2)=2(\sqrt{1+\chi^2}-1)$ \citep[see][]{Triggs2000}. This alleviated the problem by removing a few outliers from the analysis. However, a better solution for the problem is to assume that components A and B are, from the kinematics point of view, a joint component. 

As discussed in the photometric modeling (see Section~\ref{sec:sercomponents}), we performed a photometric decomposition to obtain the best description of the surface brightness in NGC~3311, but it does not necessarily follow that all these components are physically distinct. A closer inspection of the center of NGC~3311 reveals the presence of a large dust lane, and the use of two compact components (A and B) in the center may be not related to physically independent components, but just as a way to describe the central component better from a mathematical perspective. This is the case of NGC~7507, studied by \citet{2013ApJ...766...47H}, where the use of an additional component increases the quality of the photometric decomposition considerably. We believe that the center of NGC~3311 is a similar case; that is, both components A and B are used to describe the surface brightness of the same entity, as it was indicated also by a very similar kinematics derived for these components when treated disjointly.

\subsection{Results}

\begin{figure*}[t]
\centering
\includegraphics[width=\linewidth]{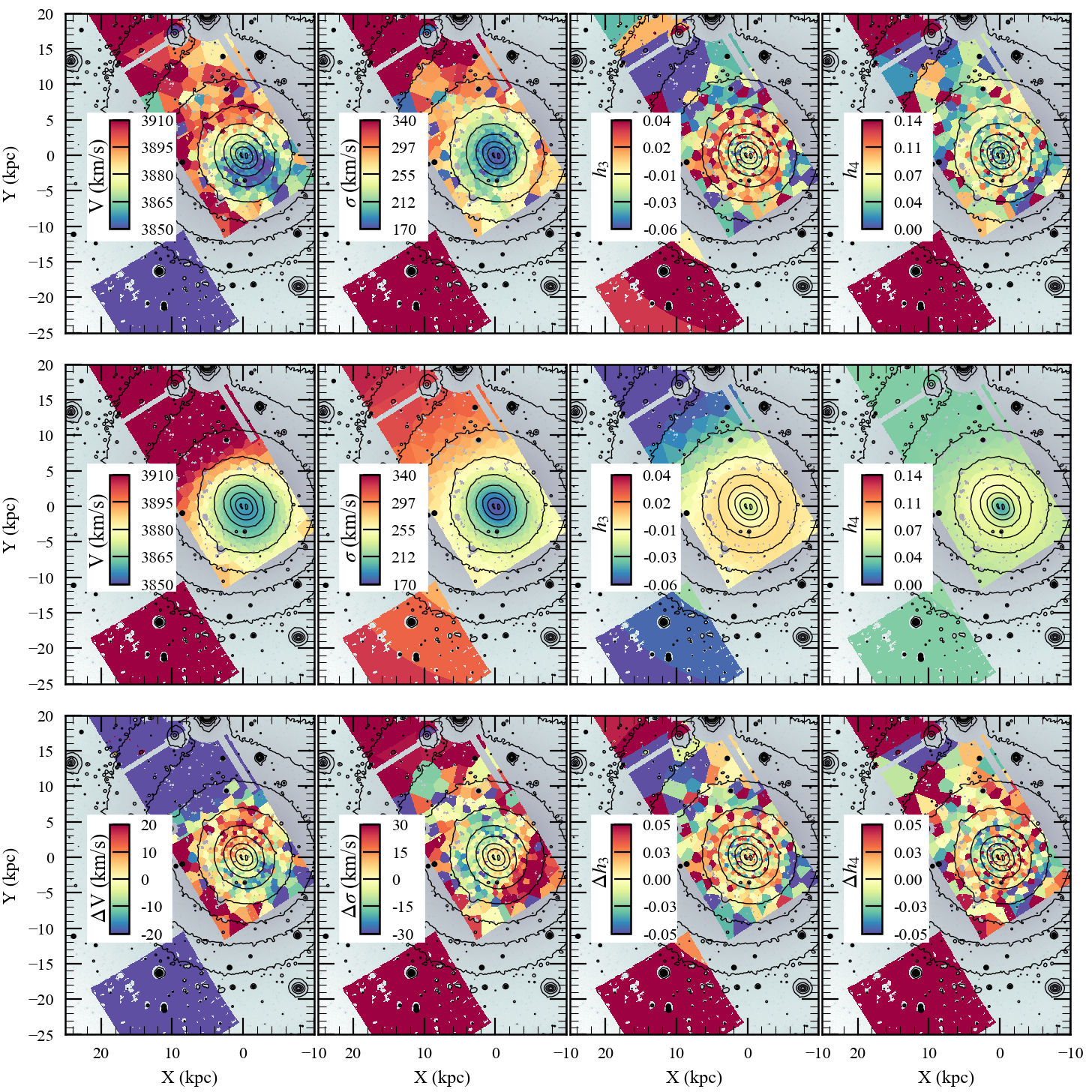}
\caption{Comparison of observed kinematics and the finite mixture model in NGC 3311. Top panels: Observed four moments of the \ac{LOSVD} in a Gauss-Hermite expansion, including systemic velocity ($V$), velocity dispersion ($\sigma$), and two high-order moments for the skewness ($h_3$) and kurtosis ($h_4$) of the distribution. Middle panels: Best fit for the finite mixture model according to parameters in Table \ref{tab:gmm} (see text for details). Bottom panels: Residuals between the observed fields on top and the best fit models in the middle panels.}
\label{fig:photkinmodel}
\end{figure*}

\begin{figure}
\centering
\includegraphics[width=\linewidth]{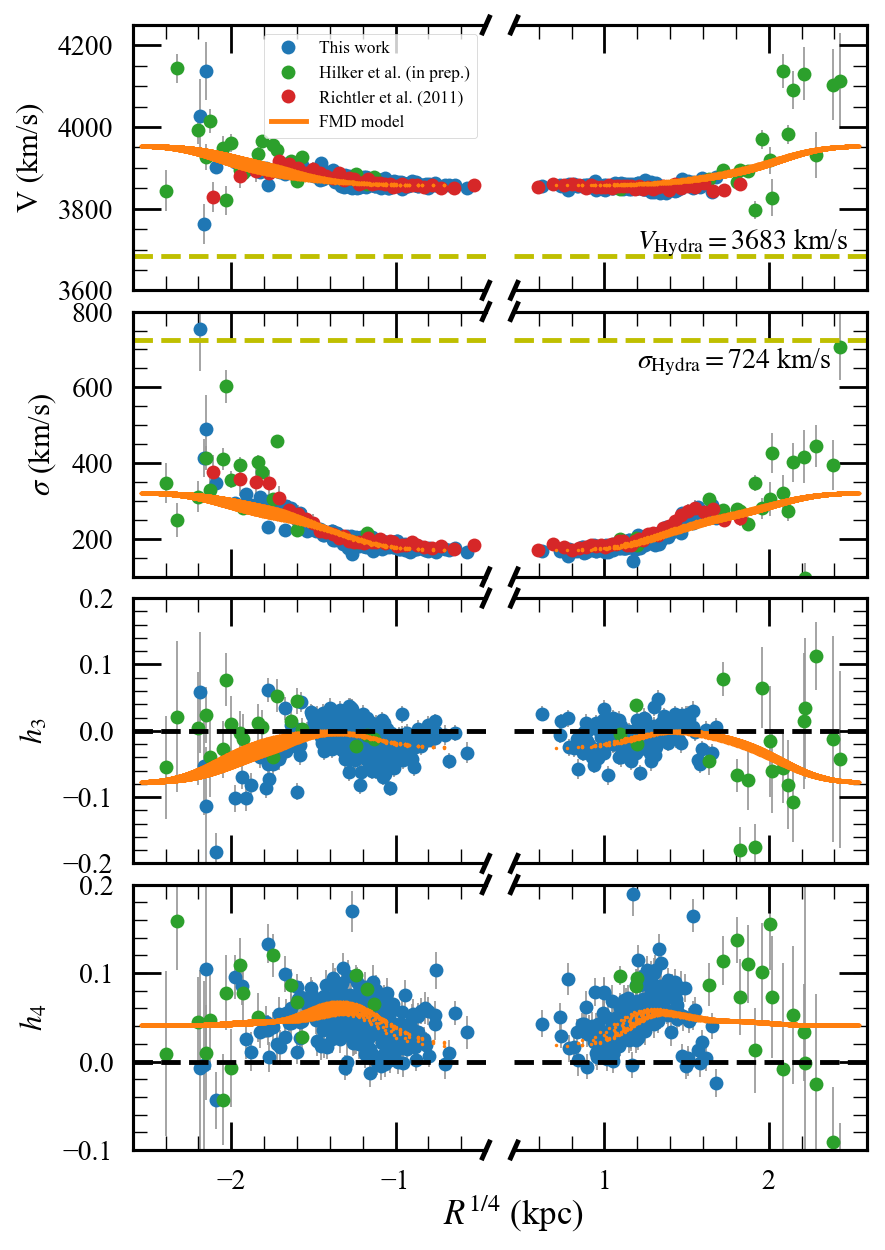}
\caption{Comparison of the radial profiles of the four moments of the \ac{LOSVD} observed along the semi-major axis at PA$=$\SI{40}{\degree} within a pseudo-slit of \SI{3}{\arcsec} width. Large blue circles indicate the observed values while small orange circles indicate the results for the finite mixture model. Negative (positive) radius indicate data points at the NE (SW) from the center of NGC 3311.}
\label{fig:kinprofile}
\end{figure}

The parameters of our best model are presented in Table~\ref{tab:gmm}. As discussed above, components A and B are treated with a unique \ac{LOSVD}, thus we refer to these components jointly as A+B. The resulting maps of the \ac{LOSVD}s are displayed in Fig.~\ref{fig:photkinmodel}, including the observed \ac{LOSVD}s (top panels), the best fit model for the \ac{FMD} model (middle panels), and the residuals between observations and models (bottom panels). To illustrate the radial profiles for the kinematics, Fig. \ref{fig:kinprofile} shows the comparison of the observed \ac{LOSVD}s with the models along a pseudo-slit orientated at a position angle of \SI{40}{\degree}, similar to the profiles presented in Fig. \ref{fig:kinmajaxis}, but now using a logarithmic scale for the radial direction to improve the visualization of the inner data points. 

\begin{table*}
\caption{Gauss-Hermite profile parameters for the four photometric components in the finite mixture model.}
\label{tab:gmm}
\centering
\begin{tabular}{ccccc}
\hline
\hline
Component & $\mu_i$ (km / s) & $\sigma_i$ (km / s) & $h_{3,i}$ & $h_{4,i}$ \\
\hline
A+B & $3856.7 \pm 1.5$ & $152.8 \pm 2.6$ & $-0.056 \pm 0.008$ & $-0.076 \pm 0.019$ \\
C & $3836 \pm 6$ & $188 \pm 7$ & $0.000 \pm 0.020$ & $-0.010 \pm 0.027$ \\
D & $3978 \pm 13$ & $327 \pm 9$ & $-0.097 \pm 0.011$ & $0.036 \pm 0.012$ \\
\hline
\hline
\end{tabular}
\end{table*}

{\it Evaluation of the model $-$} 
The major limitation of our modeling are the results at large radii, where the model has a velocity dispersion that is usually smaller than the data. However, there are a number of reasons to explain this results. The first is the presence of substructure at large radii, which increases the scatter of the measured \ac{LOSVD}. Second, also at large radii, we are dealing with bins with low S/N and larger area in relation to the central bins, which lead to larger uncertainties and, consequently, lower weights in the optimization routine. Finally, there is the issue of the spatial coverage of our observations, which cover only part of the galaxy. Despite these issues, the model is clearly able to recover the large-scale properties of the observations, e.g., the  rising of the LOS velocity and velocity dispersion from the center of NGC~3311, and the large scale northeast and southwest asymmetry, and the residual maps indicate only features on smaller scales that our model was not meant to fit. An interpretation of the scatter to larger velocity dispersions at large radii is given in \citet{hilker2017}.

{\it Component C as a fast rotator $-$} Our \ac{FMD} modeling allows us to revisit the rotation pattern observed in MUSE Field~I previosuly discussed in Section~\ref{sec:starvel}. NGC~3311 is classified as a typical slow rotator, as evidenced by the last panel of Fig.~\ref{fig:kinemetry} where we showed that the residuals from the \textsc{kinemetry} are always larger than 5\% of the rotation velocity. However, at least in the radial range between $R\sim 2$ and \SI{5}{\kilo\parsec}, the ratio $k_5/V_{\rm rot}$ is not much larger than such threshold.  In this radial range, component C dominates the light (see Fig.~\ref{fig:lightfraction}) and may be characterised by a regular rotation pattern. 

The maximum rotation in the above radial range is about $V_{\rm C}\approx$\SI{20}{\kilo\meter\per\second}, with the value for the velocity dispersion of $\sigma_{\rm C}=$\SI{188}{\kilo\meter\per\second} according to the \ac{FMD} model. Thus $V_{\rm C}/\sigma_{\rm C}\approx 0.1$, and the ellipticity of component C from the photometry is $\varepsilon=1-\frac{b}{a}=0.03$. Given these values, it is possible to classify this component as an oblate isotropic spheroid, similar to the fast rotators in the ATLAS$^{\text 3D}$ survey \citep{2011MNRAS.414..888E}. 

While NGC~3311 as a whole is correctly characterized as slow rotator, we have evidence that an individual structural component, i.e. component C, may be regarded as rotating. The implication of the classification of component C as a fast rotator denotes the importance of a detailed analysis of the kinematics in conjunction with photometry, in particular in a complex system such as a cD galaxy, where a large fraction of the light may come from accreted stars. 

{\it Kinematical signatures of asymmetric features $-$}
Our \ac{FMD} model does not include non-axisymmetric components, such as the elongated features seen in the Southwest region of MUSE Field~I (see Fig.~\ref{fig:photres}). Yet, this substructure is clearly present, as indicated by the features in the residual map of the velocity dispersion, Fig.~\ref{fig:photkinmodel}, where a local enhancement in the velocity dispersion at around $\sim 30$ \si{\kilo\meter\per\second} is observed at the same location. The shape of this substructure is unclear considering the limited area coverage of our MUSE mosaic in that specific region.

The presence of a large number of substructures around \ac{BCG}s is expected from simulations \citep[e.g.][]{2015MNRAS.451.2703C}. Therefore, our model may be used for another application, which is to verify whether streams observed in the photometric decompositions are also imprinting any kinematic signatures. The study of this substructure is beyond the scope of this project, but the confirmation of its existence in the residuals from the model highlights the importance of using a 2D analysis to study massive \ac{ETGs}. 

\subsection{The scaling laws of the components in NGC~3311}

The velocity dispersion is considered a proxy for the galaxy mass, and  it is among the primary parameters to be used in the scaling relations, such as the Faber-Jackson relation \citep{1976ApJ...204..668F} and the fundamental plane \citep{1973A&A....23..259B,1987ApJ...313...59D,1987ApJ...313...42D} relations. Here, we investigate whether the properties of the photometric \& kinematic components in NGC~3311 comply with scaling relations for isolated ETGs. 

\begin{figure}
\centering
\includegraphics[width=\linewidth]{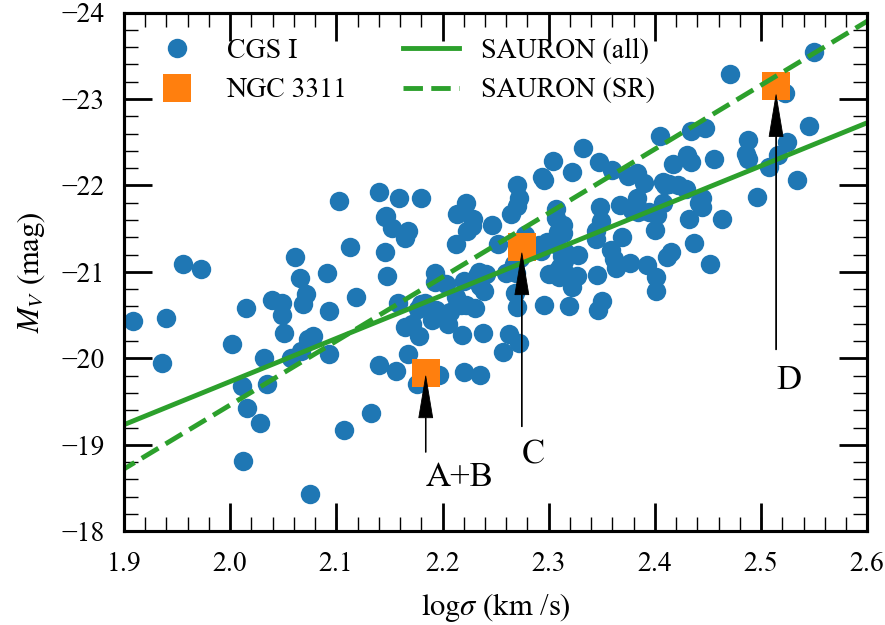}
\caption{Faber-Jackson relation for the subcomponents in NGC~3311 (orange squares) using the V-band photometric model and the velocity dispersion from the finite mixture model. Blue circles indicate the galaxies in the Carnegie-Irvine Galaxy Survey \citep{2011ApJS..197...21H}, while the green lines indicate the results from the SAURON survey \citep{2011MNRAS.417.1787F}, including all galaxies (solid line) and slow rotators (dashed line) only.}
\label{fig:faberjackson}
\end{figure}

Fig.~\ref{fig:faberjackson} shows the Faber-Jackson relation in the V-band, i.e., the total integrated magnitude as a function of the velocity dispersion. For comparison, Fig.~\ref{fig:faberjackson} also shows the Faber-Jackson relation for galaxies in the Carnegie-Irvine Galaxy Survey \citep{2011ApJS..197...21H} for a sample of 605 bright galaxies in the Southern hemisphere. We also indicate the two Faber-Jackson relations obtained by the SAURON survey \citep{2011MNRAS.417.1787F}, one including all the galaxies in the survey (solid line) and another including only slow rotators (dashed line). The central component A+B falls below the expected Faber-Jackson relation, but we may attribute that effect to the dust extinction in the central region. The analysis of the emission lines in the central region indicates the extinction is relevant within the inner kiloparsec, having values of E(B-V)$\approx 0.75$ in the center. Components C and D do follow the Faber-Jackson relations: component C lies between the relations from the SAURON survey for fast and slow rotators, and component D falls onto that for slow rotators.

\begin{figure}
\centering
\includegraphics[width=\linewidth]{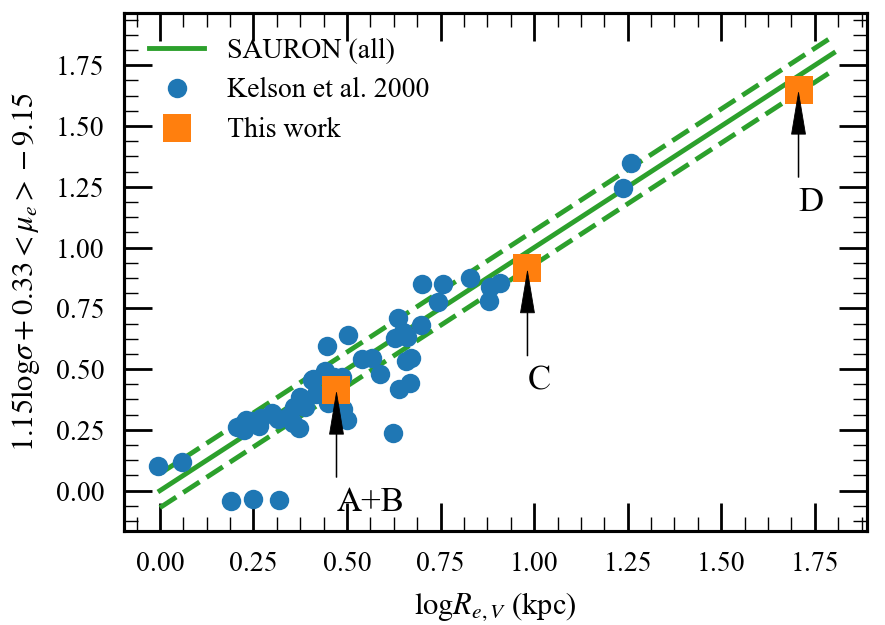}
\caption{Edge-on projection of the fundamental plane as a function of the effective radius. We use the fundamental plane equation (solid green lines) from the SAURON survey \citep{2011MNRAS.417.1787F}, offset by a factor of 0.34 to match the observations from \citet{2000ApJ...531..184K} (blue circles). The dashed green lines indicate the one standard deviation of the fundamental plane. The components of NGC~3311 identified in our analysis (orange squares) are labeled according to Table \ref{tab:gmm}.}
\label{fig:fundamentalplane}
\end{figure}

In Fig.~\ref{fig:fundamentalplane}, we show the edge-on projection of the fundamental plane for the three components of NGC~3311. For comparison, we also show the data from \citet{2000ApJ...531..184K}, who studied a sample of 53 galaxies in the cluster  Cl 1358+62 at $z=0.33$ to construct the fundamental plane for 30 elliptical and lenticular galaxies. The y-axis shows the fundamental plane projection calculated for the SAURON survey by \citet{2011MNRAS.417.1787F} for the V-band. To match the data from the SAURON with the data from \citet{2000ApJ...531..184K}, we offset the fundamental plane by 0.34. Such an offset may be caused by the different methods used for the calculation of parameters: \citet{2011MNRAS.417.1787F} use IFU data to calculate luminosity-weighted parameters within one effective radius, while \citet{2000ApJ...531..184K} use total magnitudes taken from \ac{HST} WFPC2 imaging translated to Johnson V magnitudes, whose photometry was obtained by modeling the surface brightnesses with $R^{1/4}$ profiles and central velocity dispersions \citep[see details in ][]{2000ApJ...531..137K,2000ApJ...531..159K}. We converted their effective radius from arcsec to kpc assuming distances calculated from the redshift of the cluster in the cosmology assumed in this work.

The tightness of the fundamental plane is regarded as a degree of the virialization of galaxies \citep{1987nngp.proc..175F}. The results of our analysis indicate that the components follow the expected fundamental plane relation for galaxies, including the most luminous components C and D, whose effective radii are larger than those sampled in both SAURON and  \citet{2000ApJ...531..184K} samples. The values of $\log R_{e,V}$ and $\log \sigma$ for component D place it at the high effective radius and sigma end of the fundamental plane for isolated ETGs. This region of the fundamental plane occurs before significant changes in the M/L ratios take place, such as is found in cases with ICL spheroids \citep{2006ApJ...638..725Z}.

\section{Discussion}
\label{sec:discussion}

The \ac{MUSE} IFU enables measurements of the spectra from the stellar light in the faint outer regions of ETGs. By coadding field portions it is possible to achieve adequate signal-to-noise ratios and reach out to several effective radii. Such an observing strategy is particularly well suited for \ac{BCG}s, as it enables measurements of the 2D kinematics maps and metallicities out to very large radii, sampling the extended halos of these objects \citep{2016MNRAS.461..230E}. Our goal for NGC~3311 is to measure the \ac{LOS} velocity offsets between the galaxy's bright central regions and the extended halo directly from the Doppler shifts of the absorption lines, and the 2D maps of the four LOSVD moments. The next step will be the analysis of the metallicities (in a forthcoming paper, Barbosa et al. 2018, in prep.).

\subsection{The peculiar velocity of NGC~3311}
\label{sec:pecvel}
We start our discussion with the results for the mean \ac{LOS} velocities of the different regions of NGC~3311 obtained by our 2D modeling of the velocity field.  Our mean \ac{LOS} velocity for the very center of NGC~3311, $v_{LOS,cen} = 3856\pm 10$ \si{\kilo\meter\per\second}, agrees with the previous measurements of \citet{2008A&A...486..697M, 2010A&A...520L...9V, 2011A&A...531A.119R}\footnote{See extensive discussion in \citet{2010A&A...520L...9V} on the origin of the systematic offsets with respect to the \ac{LOS} velocity measurements of \citet{2008MNRAS.391.1009L}.}.  We can compare this value with that of the mean Hydra~I cluster redshift from extended spectroscopic surveys of cluster galaxies \citep{1994AJ....107.1637B, 2003ApJ...591..764C}. This gives $\langle V_{\rm Hydra} \rangle = 3683 \pm 46$ \si{\kilo\meter\per\second} \citep{2003ApJ...591..764C}, indicating a relative motion of $\Delta V_{LOS} = 173$ \si{\kilo\meter\per\second} for the center of NGC~3311. 


Our new MUSE data cubes provide direct evidence for the galaxy's peculiar velocity from the wavelength shifts of the faint absorption lines on the continuum from the inner bright regions to the faint outer  halo.
The visual inspection of the mean \ac{LOS} velocity map showed already that the inner regions of the galaxy, corresponding to photometric and kinematic components A to C, have a different mean velocity from the diffuse outer envelope, the D component. The results of the \ac{FMD} modeling indicates that the D component is offset by $\Delta V_{LOS}=122\pm 13$ \si{\kilo\meter\per\second} and $\Delta V_{LOS}=142\pm 14$ \si{\kilo\meter\per\second} from component A+B and C, respectively.  Assuming a cluster velocity dispersion from satellite members of $\sigma_{\rm Hydra}=724\pm 31$ \si{\kilo\meter\per\second} \citep{2003ApJ...591..764C}, our modeling of the MUSE 2D \ac{LOS} velocity map indicates that component D, i.e. the envelope of NGC~3311, has a peculiar LOS velocity of at least $20\%$ of the cluster velocity dispersion with respect to the inner galaxy. These results greatly improve on the early attempts by  \citet{1994AJ....107.1637B} to measure the peculiar velocities of this \ac{BCG}.

\subsection{The cD envelope of NGC~3311}
\label{sec:cDenv}
In the local universe, peculiar velocities of \ac{BCG}s with respect to their diffuse envelope or the intracluster light component have been measured for M87 in Virgo  \citep{2015A&A...579A.135L}, NGC~6166 in Abell 2199 \citep{2015ApJ...807...56B} and NGC~4874 in the Coma cluster \citep{2007A&A...468..815G}.  In NGC~3311, the outer envelope (the D component) dominates over the entire light distribution of NGC~3311, providing 81\% of the total light. This outer envelope is not only shifted relative to the central galaxy in the mean velocity, by $\Delta V_{LOS}=142\pm14 \si{\kilo\meter\per\second}$ as discussed in Section \ref{sec:pecvel}, but also in space, by about $33''$ (8.6 kpc, this work) to $50''$ \citep[12.5 kpc;][]{2012A&A...545A..37A}. Thus also a tangential relative motion must be present, so that we estimate a total relative velocity between central galaxy and cD envelope, of $\Delta V_{tot} = \sqrt{3}\times\Delta V_{LOS} = 204$ \si{\kilo\meter\per\second}. 

If we follow the definition of the velocity bias of a \ac{BCG} given in \citet{2005MNRAS.361.1203V}, and adopt  $\langle\sigma_{cen}\rangle = \Delta V_{tot}$ and $\langle\sigma_{\rm sat}\rangle = \sigma_{\rm Hydra}$, then the total velocity bias of component D is about $b_{vel}=0.3$. We can also compare the spatial offset measured for component D with the estimates from the models of galaxy velocity biases from \citet{2005MNRAS.361.1203V} and \citet{2015MNRAS.446..578G}. These works agree on an estimate of the radial bias $b_{rad} \simeq 1\% $ of the halo virial radius. For halo virial masses $ \geq 10^{14} h^{-1} M_\odot$, $R_{vir} \sim 1$ Mpc and the radial bias value is thus $\simeq 10$ kpc, which is of the same order of magnitude as the spatial offsets measured for component D with respect to the NGC~3311 central regions.

Within the paradigm of galaxy formation in $\Lambda$CDM cosmology, \ac{BCG}s are expected to have non-negligible peculiar velocities in relation to the cluster dark matter halo \citep{2005MNRAS.361.1203V,2017ApJ...841...45Y}. Because of the large velocity dispersion and spatial extension, the cD envelope lives in a larger volume and its stars move in the potential generated primarily by the closest dark matter distribution, in the cluster core \citep{2010MNRAS.405.1544D}. Thus one might expect that the cD envelope is closer in velocity to the cluster core, and that the central galaxy would have been perturbed out of the center of this larger structure, e.g., because of close interactions with other satellite galaxies \citep{2007MNRAS.377....2M}. This case is dubbed {\it Non-Relaxed Galaxy} (NRG) scenario in the study of velocity bias by \citet{2005MNRAS.361.1203V} and \citet{2017ApJ...841...45Y}. However, the peculiar velocity of a \ac{BCG}s may also arise when the innermost parts of the cluster halo are in motion relative to the cluster dark halo on larger scales. This case is dubbed the {\it Non-Relaxed Halo} (NRH) scenario  \citep{2005MNRAS.361.1203V}. NRHs may occur because of subcluster mergers, that could be recognized because they also leave imprints in the X-ray emission and temperature maps \citep{2004MNRAS.352..508R}.   

Which of these scenarios applies to NGC 3311? 
From \citet{2006PASJ...58..695H, 2012A&A...545A..37A} and our current study, we know that the D component is co-spatial with a high metallicity region in the X-rays $\sim 1.5$ arcmin north-east of NGC~3311. It also coincides spatially with the X-ray emission from the Hydra~I cluster core: see the X-ray image in Fig.~2 of \citet{2004PASJ...56..743H}. A faint extended X-ray emission with angular dimension $\sim 1'$ trailing NGC~3311 to the northeast towards this high metallicity region is also seen. This could be due to ram pressure stripping of the gas  \citep[see Chandra observations from][]{2004PASJ...56..743H}, if this feature and the confined ($<2$kpc) strongest X-ray emitting region to the south-west, which is co-spatial with the optical inner region of NGC~3311, are moving with relative velocity $\simeq 260$ \si{\kilo\meter\per\second} during the last $5 \times 10^7$ yr \citep{2004PASJ...56..743H,2006PASJ...58..695H}. This relative velocity is consistent with the velocity of component D relative to A+B \& C.
If component A+B \& C are moving with respect to D, we might then expect the outer envelope to be at rest with respect to the mean cluster redshift, as observed for example for NGC~6166 in the Abell~2199 cluster \citep{2015ApJ...807...56B}.

What is the {\it total} relative velocity of component D with respect to the mean cluster redshift of the core region and the cluster on large scales? The mean \ac{LOS} velocity of the galaxies in the core is $3982 \pm 148$ \si{\kilo\meter\per\second} \citep{2008A&A...486..697M} and agrees with the mean velocity of the D component, within the error. The difference with the mean cluster redshift $\langle V_{\rm Hydra}\rangle$ is $\Delta V_{LOS} = 295$ \si{\kilo\meter\per\second}. If we account for the tangential relative motion associated with the spatial offset between the X-ray high metallicity peak and the center of the ICM emission of the Hydra~I cluster, which is $28''$ to the south west of NGC~3311 \citep{2004PASJ...56..743H}, we get $\Delta V_{tot} = \sqrt{3}\times 295 = 510 $ \si{\kilo\meter\per\second}. Hence the inner core halo and the envelope have a velocity bias of at least 40\% of the cluster velocity dispersion.  The large value of the velocity bias of the cD envelope and the cluster core is similar to what is found for \ac{BCG}s in Abell clusters \citep{2014ApJ...797...82L}. The evidence from the X-ray distortions and the large velocity bias of the envelope support the NRH scenario for the peculiar velocities of the different components (A+B, C, and D) in NGC~3311.


What is causing the sloshing of the inner core in the Hydra~I cluster? Within the inner $100 \times 100$ kpc$^2$ region centred on NGC~3311, there are 14 galaxies in total with $V>4450$  \si{\kilo\meter\per\second} that are currently being disrupted and adding debris to the outer envelope \citep{2011A&A...528A..24V, 2012A&A...545A..37A}.  On a slightly larger scale, i.e. 15 arcmin ($\simeq 236$ kpc radius), the mean average redhifts of satellite galaxies is $3982 \pm 148$ \si{\kilo\meter\per\second} \citep{2008A&A...486..697M} that also is offset to a larger velocity from the value determined by \citeauthor{2003ApJ...591..764C} over the entire cluster. This is suggestive of an on-going subcluster merging. In such case, the peculiar velocities of the NGC~3311 component D and A+B,C can be understood as the central regions being pulled along by the portion of the inner cluster halo closer to the envelope, because of the on-going subcluster merger. Since the inner high density regions experience stronger deflections \citep{2007A&A...468..815G,2007MNRAS.377....2M,2011A&A...528A..24V}, and because of different projections of component  velocities relative to the LOS,  the LOS velocities with respect to the cluster at larger scales  may be  different for component D compared to components A+B and C.

\subsection{High-order kinematics moments}
\label{sec:highorder}

Ordinary massive ETGs have nearly isothermal mass profiles and nearly Gaussian LOSVDs, with small values of the Hermite higher moments ($h_3$, $h_4$) indicating slow rotation and modest radial anisotropies \citep[e.g.][]{2001AJ....121.1936G, 2013ApJ...777...98S, 2015ApJ...804L..21C}. Generally the deviations from Gaussian LOSVDs are small, with the asymmetric deviations signalled by the $h_3$ parameter values being usually larger than the symmetric $h_4$  values \citep{1994MNRAS.269..785B}. Our \ac{MUSE} 2D maps for the moments of the LOSVD in NGC~3311 show relatively large spatial variations for both $h_3$ and $h_4$, with positive values of $h_4$ larger than the $h_3$ values over the entire area.   
Our model in Sec.~\ref{sec:modeling} indicates that most of the spatial variations of the $h_3$ and $h_4$ maps can be reproduced by the superposition of different components along the \ac{LOS} whose higher order moments (i.e. the deviations from Gaussian LOSVD) are small. It is the superposition of nearly isothermal spheroids with different light distributions, LOS velocities and centers that causes the spatial variations in the 2D maps and profiles as seen in Fig.~\ref{fig:photkinmodel} and \ref{fig:kinprofile}. From the \ac{FMD} modeling, the inner regions, i.e. components A+B \& C, are characterised by  small negative (-0.056) or null values of the $h_3$ parameter and small negative values (-0.076, -0.010 respectively) for $h_4$, while the outer envelope is associated with a modest negative value (-0.097) for $h_3$ and positive value (0.036) for $h_4$. The superposition of components along the LOS for NGC~3311 reproduces the $h_3,\, h_4$ vs $v/\sigma$ correlations shown in Fig.~\ref{fig:h3h4}. These correlations are similar to those computed for simulated galaxies formed from dry mergers or already passive progenitors, according to the {\it F type} classification in \citet{2014MNRAS.444.3357N}.

Recent studies based on extended IFU, e.g. the MASSIVE survey \citep{2017MNRAS.464..356V}, and deep long slit \citep{2015ApJ...807...56B} observations measured the values of $h_3$, $h_4$ out to two effective radii and showed that the most luminous ETGs \citep[$M_K < -26.0$;][]{2017MNRAS.464..356V} have mean values of $\langle h_4\rangle \simeq 0.05$ at $2R_e$ and close to zero in the centers \citep[e.g.][]{2017ApJ...835..104V, 2017MNRAS.464..356V}.
For NGC~3311, the $h_3,\, h_4$ profiles are shown in the two bottom panels of Fig.~\ref{fig:kinprofile}.  $h_3$ is nearly zero out to $2$ kpc and then becomes negative at distances larger than $10$ kpc, where the \ac{LOS} velocity increases. $h_4$ values are near zero at the center, increase to $\simeq 0.1$ at $2-4$ kpc and then decrease to smaller positive values at radii $>10$ kpc. The $h_4$ positive values measured for NGC~3311 are similar to those measured in other galaxies of the same luminosity class \citep{2017MNRAS.464..356V}.  

The core-wing structure of the LOSVD that is signalled by positive values of $h_4$ may come about because of radial anisotropy, that causes an overabundance of stars at zero \ac{LOS} velocity \citep{1998MNRAS.295..197G}. Radial anisotropy causes the projected LOS dispersion to be underestimated also, and leads to a decreasing $\sigma_{LOS}$ profile with radius \citep{2007ApJ...664..257D, 2009MNRAS.395...76D}. Positive $h_4$ and {\it increasing $\sigma$} as observed in NGC~3311 are reminiscent of the properties of simulated massive ETGs in cosmological simulations by \citet{2014MNRAS.438.2701W}. These most massive galaxies  have increasing circular velocity curves with radius, radial anisotropy in the $ 2 - 5 R_e$ region and a large fraction of accreted stars \citep{2014MNRAS.438.2701W}.

In Sec.~\ref{sec:modeling}, the results from our modeling show that each component has a modest or small anisotropy, with the inner regions, i.e. components A+B \& C, characterised by negative values for $h_4$ as in the case of tangential anisotropy.
The \citet{2014MNRAS.438.2701W} simulations predict similar values of $h_4$ for components created by dissipative events. Because the outermost component (component D) has a surface brightness profile with the largest effective radius ($R_{e,D}=51\pm5$ kpc) and velocity dispersion $\sigma_D=327$ \si{\kilo\meter\per\second}, it contributes light in the inner $10$ kpc from regions with larger circular velocity, giving thus rise to the core-winged structure of the LOSVD, as it is measured.



\section{Summary and conclusion}
\label{sec:conclusion}

In this work, we presented the detailed analysis of the MUSE mosaiced pointing covering the light of NGC~3311, that allowed the detailed mapping of the \ac{LOSVD} moments within the central galaxy and part of its extended cD halo. In the first part of the paper, we measured the LOSVD maps from our new MUSE data. Based on the extended maps, we determine  the peculiar velocity of the central galaxy in relation to its stellar halo, the rising velocity dispersion profile, and the presence of a cD halo that is kinematically decoupled from the central luminous regions of NGC~3311. The entire galaxy is classified as a slow rotator, with  small correlation between the high-order moments $h_3, \, h_4$ with $V/\sigma$. The diagnostic diagrams $h_3$, $h_4$ vs. $V/\sigma$ indicate that NGC~3311 is similar to simulated galaxies formed mostly by dry mergers of already passive progenitors.

In the second part of the paper, we revisited the photometry of NGC~3311. Using deep V-band photometry and building upon previous results on the surface brightness decomposition, we carried out a photometric model that includes four S\'ersic components. Three out of four components are concentric, while the outer one has a central offset and dominates the light at large radii. This model is able to reproduce the surface brightness distribution in the galaxy and its immediate surroundings in the Hydra~I cluster core. The residuals further indicated the presence of asymmetric substructures pointing to recent episodes of minor accretion events.  

In the final part of this paper, we developed a simple finite mixture model to describe the \ac{LOSVD} of the system based on the photometric subcomponents and assuming nearly isothermal spheres with small deviations from Gaussianity. This model describes the large-scale properties of the observations very well, including the strong radial variations of the four \ac{LOSVD} moments and their 2D maps.  We conclude that the outer envelope is responsible for the core-winged structure of the LOSVD in the bright central region also, by contributing stars that experience larger circular velocities. The success of this modeling supports the combined approach, surface brightenss profiles and 2D kinematics modeling, to study massive, non-rotating early-type galaxies, and clarify the origin of the radial gradients in the four \ac{LOSVD} moments in \ac{BCG}s. 

On the basis of these results, we present a strong case for the peculiar velocity of the outer halo of NGC~3311 and measured a velocity bias of 0.3 for this \ac{BCG}, as a lower limit. The results of the FMD modeling show that the envelope (component D) i) is displaced from the central regions, and ii) has a larger effective radius, and iii) larger mean velocity and velocity dispersion compared to the central regions. These results, taken together with the morphology of the cospatial X-ray emission, let us conclude that the velocity bias originates from a ``Non Relaxed Halo'' scenario caused by an on-going subcluster merging in the Hydra~I core. We conclude that the cD envelope of NGC~3311 is dynamically associated to the cluster core, which in Hydra~I cluster is in addition displaced from the main cluster center, presumably due to a recent subcluster merger.

The result of our analysis depicts NGC~3311 as a massive elliptical whose majority of stars are accreted from merging satellities orbiting in the Hydra~I cluster, on preferentially radial orbits. Our next steps will be to constrain the progenitors of the components in NGC~3311, by measuring age and metallicity of their stars.

\begin{acknowledgements}
Based on observations collected at the European Organisation for Astronomical Research in the Southern Hemisphere under ESO programmes 065.N-0459(A), 088.B-0448(B) and 094.B-0711(A). CEB and CMdO thanks the S\~{a}o Paulo Research Foundation (FAPESP) funding (grants 2009/54202-8, 2011/21325-0, 2016/17119-9 and 2016/12331-0). MAR and CEB acknowledge support from the ESO DG discretionary funds. TR acknowledges support from the BASAL Center for Astrophysics and Associated Technologies (PFB-06/2007). TR also acknowledges an ESO senior visitorship during May/August 2016.
\end{acknowledgements}

\bibliographystyle{aa} 
\bibliography{biblio}

\appendix

\section{Effect of the interstellar medium absorption lines to the LOSVD}
\label{sec:ism}

Absorption lines from the \ac{ISM}, e.g. Fraunhofer D-lines ($\lambda 5890, 5896$), may imprint the continuum from the stars and affect the \ac{LOSVD} measured with \textsc{pPXF}. This effect was shown by \citet{2015ApJ...807...56B} in the cD galaxy NGC 6166, where the velocity dispersion profile remained constant for the sodium line, but was otherwise increasing. However, measuring the properties of these absorption lines is complicated because they are embedded into other absorption lines. While accessing the wavelength range for our analysis, we have tested the profiles for the \ac{LOSVD} parameters on top of the main optical absorption lines using the bands of the Lick indices \citep{1977AJ.....82..941F}. 

\begin{figure}
\centering
\includegraphics[width=\linewidth]{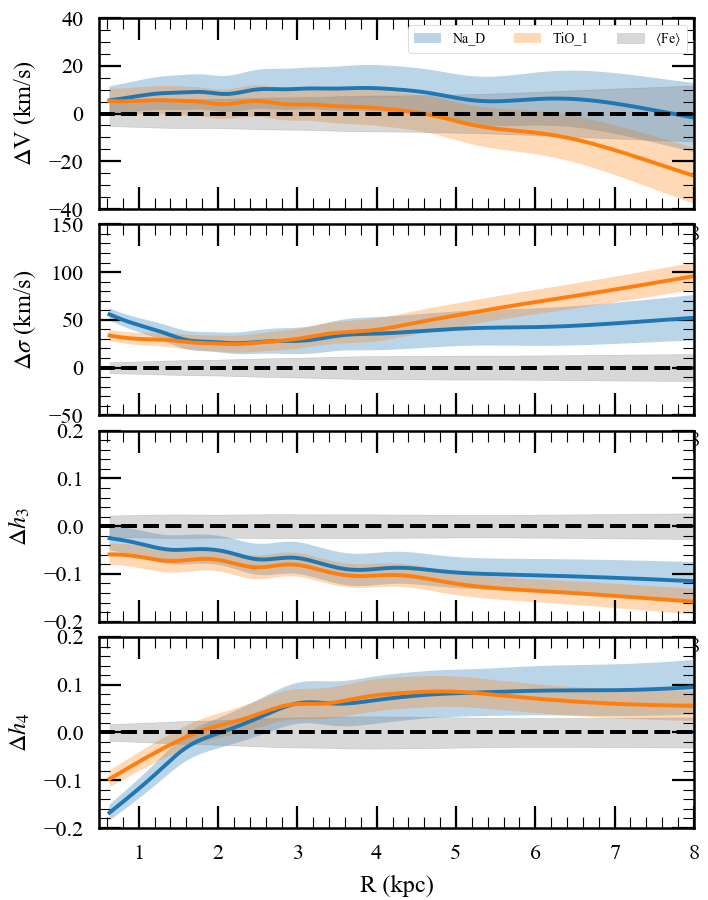}
\caption{Comparison of the LOSVD moments as a function of radius for the wavelength ranges defined according to the Lick indices in the observed field A. Solid lines display the running average of the measured values, and the shaded areas around the lines indicate the typical standard deviation from the mean. }
\label{fig:losvdlick}
\end{figure}

The absorption line features of the Lick indices are relatively narrow for the determination of \ac{LOSVD} with \textsc{pPXF} in our observational setup, so we fitted the \ac{LOSVD}s of the Lick indices considering also their blue and red sidebands according to the definitions from \citet{1997ApJS..111..377W}. The determination of the \ac{LOSVD} in these delimited wavelength increases the scatter of the profiles considerably, but we noticed that profiles for the \ac{LOSVD} parameters are compatible for almost all the indices with the exception of Na D and TiO$_{1}$, which indicate some systematic deviations. Fig. \ref{fig:losvdlick} shows the difference in the profiles of these two indices in relation to the mean profiles of the iron indices. 

The presence of the \ac{ISM} absorption D-lines may have caused both the systematically larger velocity dispersion ( by about $\Delta \sigma \simeq 20-80$ kms$^{-1}$) at the outer radius of 8~kpc and also the larger velocity dispersion value (by about $\Delta \sigma \simeq 20 - 45$ kms$^{-1}$) at $R\lesssim 1.5$ kpc. The skewness ($h_3$) values for these two absorption features is also systematically lower than the values measured for the other Lick indices. The $h_4$ profiles for the Na~D and TiO~1 absorption features show negative values at the center and a steep gradient between 1 and 2 kpc, reaching positive values of $h_4 \simeq 0.1$ at radii larger than 2.5 kpc. 

\end{document}